\documentclass[journal]{IEEEtran}
\usepackage{amsmath,amssymb}
\usepackage{subfigure}
\usepackage{graphicx,graphics,color,psfrag}
\usepackage{cite,balance}
\usepackage{caption}
\allowdisplaybreaks
\usepackage{algorithm}
\usepackage{algorithmic}
\usepackage{accents}
\usepackage{amsthm}
\usepackage{bm}
\usepackage{url}
\usepackage[english]{babel}
\usepackage{multirow}
\usepackage{enumerate}
\usepackage{cases}
\usepackage{stfloats}
\usepackage{dsfont}
\usepackage{color,soul}
\usepackage{amsfonts}
\usepackage{cite,graphicx,amsmath,amssymb}
\usepackage{subfigure}
\usepackage{fancyhdr}
\usepackage{hhline}
\usepackage{graphicx,graphics}
\usepackage{array,color}
\usepackage{mathtools}
\usepackage{amsmath}

\newtheorem{proposition}{\emph{\underline{Proposition}}}

\newtheorem{example}{\bf Example}

\def\rank{{\operatorname{rank}}}

\def\l{\left}
\def\r{\right}
\def\({\left(}
\def\){\right)}

\def\b0{{\mathbf{0}}}

\newcommand{\tr}{\mathrm{tr}}
\newcommand{\diag}{\mathrm{diag}}

\newcommand{\nn}{\nonumber}

\begin{document}
\captionsetup[figure]{name={Fig.}}
\title{\huge 
{\color{black} Channel Estimation and Passive Beamforming for Intelligent Reflecting Surface: Discrete Phase Shift \\ and Progressive Refinement}}
\author{Changsheng You,~\IEEEmembership{Member,~IEEE}, Beixiong Zheng,~\IEEEmembership{Member,~IEEE},\\
	 and Rui Zhang,~\IEEEmembership{Fellow,~IEEE}   \thanks{\noindent Part of this work will be presented at the  IEEE International Conference on Communications (ICC), Dublin, Ireland, June 2020 \cite{you2019intelligent}. 

The authors are  with the Department of Electrical and Computer Engineering, National University of Singapore, Singapore (Email: \{eleyouc, elezbe, elezhang\}@nus.edu.sg).
}}

\maketitle


\begin{abstract}
Prior studies on Intelligent Reflecting Surface (IRS) have mostly assumed perfect channel state information (CSI) available for designing the IRS passive beamforming as well as the continuously adjustable phase shift at each of its reflecting elements, which, however, have simplified  two challenging issues  for implementing IRS in practice, namely, its channel estimation and passive beamforming designs  both under the constraint of  {\it discrete} phase shifts.  
{\color{black}To address them, we consider in this paper an IRS-aided single-user communication  system with discrete phase shifts and design the IRS training reflection matrix for channel estimation as well as the passive beamforming for data transmission, both subject to the constraint of discrete phase shifts. We show that the training reflection matrix design for discrete phase shifts greatly differs from that for continuous phase shifts, and thus the corresponding passive beamforming should be optimized by taking into account the {\it correlated} channel estimation error due to discrete phase shifts. Specifically, we consider a practical block-based transmission, where each block has a finite (insufficient) number  of training symbols for channel estimation.  A novel {\it hierarchical} training reflection design is proposed to {\it progressively} estimate IRS elements' channels over multiple blocks by exploiting IRS-elements grouping and partition. Based on the resolved  IRS channels in each block, we further design the progressive passive beamforming at the IRS with discrete phase shifts to improve the achievable rate for data transmission over the blocks.}
Moreover, extensive numerical results are presented which show significant performance  improvement of the  proposed  channel estimation and passive beamforming designs as compared to various benchmark schemes.
\vspace{-5pt}
\end{abstract}
\begin{IEEEkeywords}
Intelligent reflecting surface,  channel estimation, passive beamforming, discrete phase shift.
\end{IEEEkeywords}
%
\vspace{-7pt}
\section{Introduction}\label{Sec:Intro}
The last decade has  witnessed a proliferation of innovations for wireless communications to meet its explosive growth of data traffic and ever-increasing demand for higher data rates, such as massive multiple-input-and-multiple-output (MIMO), millimeter wave  (mmWave)  communication, and so on. Although these technologies can significantly improve the spectral efficiency of wireless communication systems, they also face challenges due to the increasingly   higher hardware cost and energy consumption, which, if not successfully circumvented, may severely hinder their future applications.  Recently, \emph{intelligent reflecting surface} (IRS) and its various counterparts (such as  reconfigurable intelligent surface (RIS) and so on)  have emerged as a new and cost-effective solution to tackle these challenges \cite{qingqing2019towards,di2019smart,basar2019wireless,liaskos2018new}. Generally speaking, IRS is one kind of meta-surface composed of  a vast number of passive reflecting elements, which can be controlled in real time to dynamically alter the amplitude and/or phase of the reflected signal, thus collaboratively enabling \emph{smart reconfiguration} of the radio propagation  environment. Besides, IRS does not require any active radio frequency (RF) chains for signal transmission/reception but simply relies on passive signal reflection, thus significantly reducing the hardware cost and  energy consumption as compared to traditional active transceivers/relays. Moreover, IRS	can be easily attached to or removed from different objects (e.g., walls and ceilings), hence exhibiting great  flexibility and compatibility in practical deployment.

Despite the above appealing advantages, one critical issue in the design of IRS-aided communication systems is how to  judiciously set the reflection coefficients of its massive  elements based on the channel state information (CSI) of all signal paths, such that the signals reflected by IRS can be added constructively with those via other  paths to enhance the signal power at the intended receiver, or destructively to help mitigate co-channel interference. The design of IRS passive beamforming has been investigated in different setups, assuming continuous phase shifts \cite{wu2019intelligent,huang2019reconfigurable,zhaoMM2019,nadeem2019intelligent,wang2019joint} or discrete phase shifts \cite{wu2019beamforming,huang2018energy,guo2019weighted2} of the reflecting elements. Moreover, IRS passive beamforming has been jointly designed with other communication techniques, such as orthogonal frequency division multiplexing (OFDM) \cite{yang2019intelligent}, MIMO\cite{zhang2019capacity,pan2019multicell}, non-orthogonal multiple access (NOMA) \cite{yang2019intelligentnoma,fu2019intelligent,zheng2020intelligent1,ding2019simple}, physical-layer security\cite{cui2019secure,guan2019intelligent,chen2019intelligent,yu2019enabling}, and simultaneous wireless information and power transfer (SWIPT) \cite{wu2019joint,pan2019intelligent}, etc.

To reap the passive beamforming gain of IRS, existing works (e.g., \cite{wu2019intelligent,huang2019reconfigurable,zhang2019capacity,pan2019multicell,yang2019intelligentnoma,fu2019intelligent,mu2019exploiting,ding2019simple,cui2019secure,guan2019intelligent,chen2019intelligent,yu2019enabling}) have mostly assumed perfect CSI available for all the individual channels between the IRS and its aided access point (AP) as well as users, which, however, is practically difficult to realize due to the following  reasons. First, IRS can only reflect signals without the capabilities of signal transmission/processing, thus it is practically difficult to estimate its channels with the AP as well as users directly  \cite{qingqing2019towards}.  Instead, only the cascaded user-IRS-AP channels can be estimated at the AP (or user) based on the pilot symbols sent by the user (or AP), by adjusting IRS  reflection coefficients over time \cite{he2019cascaded,zheng2019intelligent,mishra2019channel}. Then,  IRS passive beamforming can be designed based on the estimated cascaded channels for data transmission.
Second, since IRS usually consists of a large number of reflecting elements, the conventional ``\emph{all-at-once}" channel estimation method whereby 
the cascaded channels for all IRS reflecting elements are estimated  at one time will require long pilot length that increases  with the number of reflecting elements \cite{zheng2019intelligent} and thus cause long delay for data transmission, making it unsuitable  for delay-sensitive and/or short-packet transmissions.
Moreover, the all-at-once channel estimation for IRS may be  \emph{incompatible} with the existing communication block structure, where only a small number of pilot symbols are allocated in each (time) block. 
To reduce the channel training overhead,  the IRS elements can be divided into  groups where only the effective channel for all elements in each group needs to be estimated \cite{yang2019intelligent,zheng2019intelligent}. As a result, the required number of pilot symbols  is reduced to the number of groups, instead of the number of elements in the case of all-at-once channel estimation.  However, this comes at the cost of degraded IRS passive beamforming performance since with only per-group effective channels available, the reflection coefficients need to be set identical for all elements in each group, which reduces the design degrees-of-freedom (DoF) for passive beamforming. Furthermore, the channel estimation considered in \cite{yang2019intelligent,zheng2019intelligent} based on IRS-elements grouping has assumed IRS continuous phase shifts for each of the  IRS elements, while practical   phase shifters  can only operate with a finite number of discrete phase-shift values due to the hardware constraint \cite{qingqing2019towards}. {\color{black} To summarize,  the existing works on IRS channel estimation (e.g., \cite{he2019cascaded,zheng2019intelligent,mishra2019channel,zheng2020intelligent,wang2019compressed,wang2019channel}) have mostly assumed continuous  phase shifts for IRS and exploited certain IRS channel properties (e.g., low-rank, sparsity, spatial correlation, etc.) in some specific environments for reducing the channel estimation overhead. Nevertheless, these methods are inapplicable to the IRS with practically discrete phase shifts  and  lack of  generality for arbitrary IRS channels given limited training time.
}

\begin{figure}[t]
\begin{center}
\includegraphics[height=3.7cm]{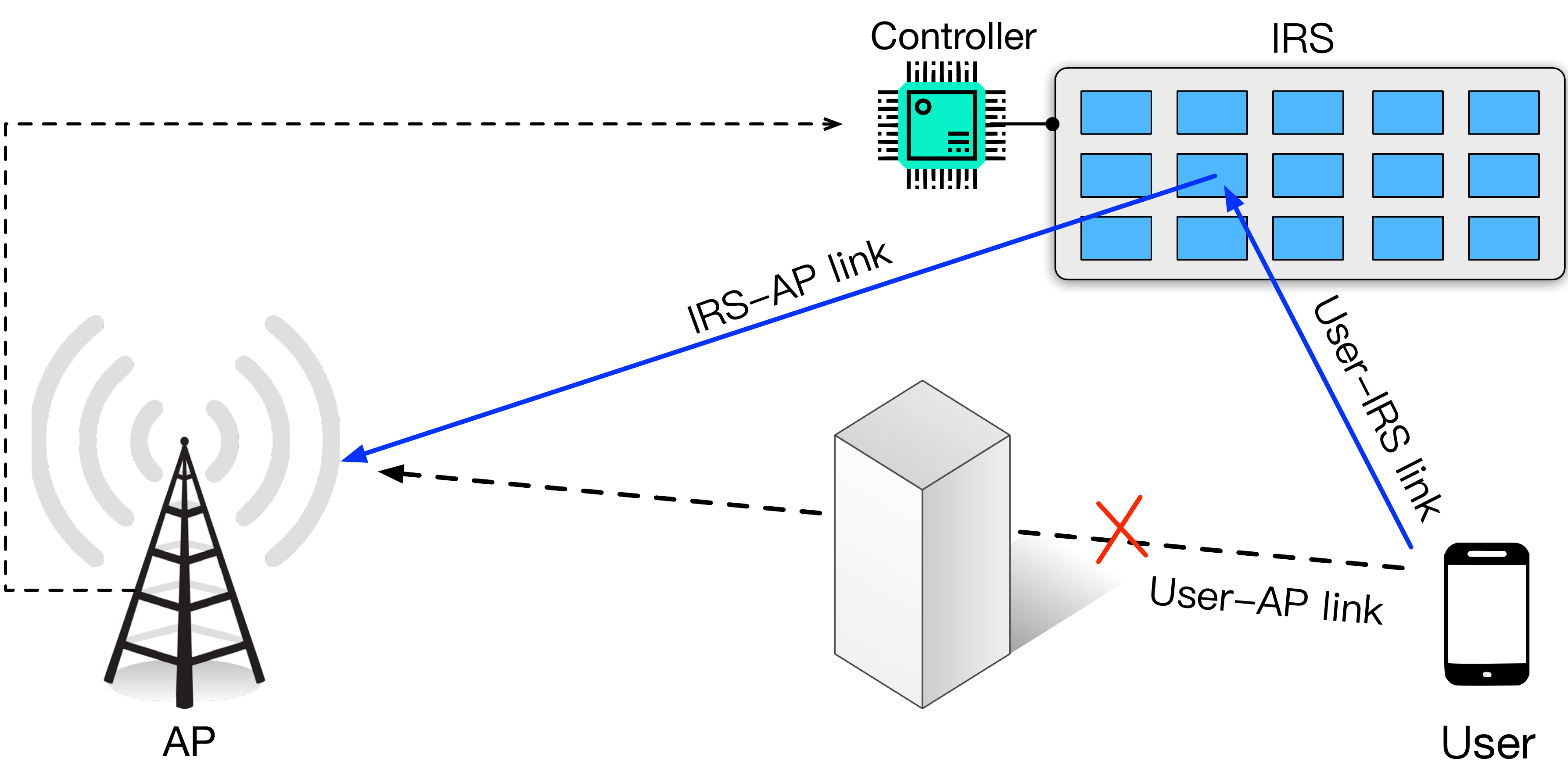}
\caption{An IRS-aided single-user communication system.}
\label{Fig:Syst}
\end{center}
\end{figure}

\begin{figure*}[t]
\begin{center}
\includegraphics[height=2cm]{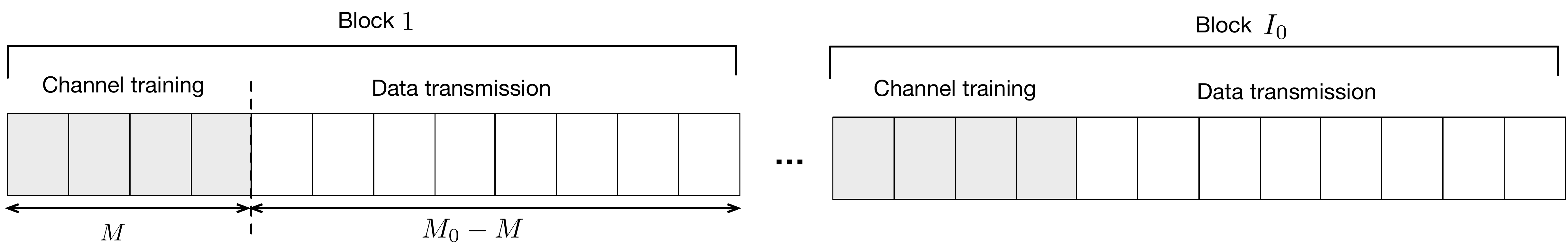}
\caption{Pilot-assisted block transmission, where the CSI is assumed constant over $I_0$ consecutive blocks.}
\label{Fig:Proto}
\end{center}
\end{figure*}

{\color{black}To overcome the aforementioned limitations in the existing studies on IRS channel estimation as well as passive beamforming and make IRS implementable in practice, we investigate their new designs in this paper by considering  the more realistic setting with  finite pilot/training  symbols in each block as well as discrete phase shifts for both IRS channel estimation and passive beamforming (for data transmission).} For the purpose of exposition, we consider an IRS-aided communication system as shown in Fig.~\ref{Fig:Syst}, where an IRS is deployed to assist the data transmission of a single-antenna user with a single-antenna AP. Based on the existing pilot-assisted block transmission (see Fig.~\ref{Fig:Proto}) in which each block consists of a finite (usually small) number of pilot symbols,    we propose a new approach to \emph{progressively} resolve the IRS elements' (cascaded) channels over the blocks and accordingly refine the IRS passive beamforming to improve the achievable rate for data transmission, by (non-trivially) extending the IRS-elements grouping method in \cite{yang2019intelligent,zheng2019intelligent}. 
The main contributions of this paper are summarized as follows.

\begin{itemize}
\item We propose a novel \emph{hierarchical training reflection design} to progressively estimate the IRS elements' channels over consecutive blocks\footnote{We assume that the CSI of all channels is unchanged over these  blocks, which is practically valid as IRS is at a fixed location and the IRS-served users are typically in short distance from IRS and of low mobility.}. The key idea is to decompose each training reflection vector, which consists of the reflection coefficients of all IRS elements in a given pilot symbol duration, into the Kronecker product of two vectors, called the (group-wise) \emph{basis training reflection} vector and the \emph{intra-group training reflection} vector (common for all groups), respectively. More specifically, the basis training reflection vectors over all pilot symbols in each block (collectively called the \emph{basis training reflection matrix} which is the same for all the blocks shown in Fig.~\ref{Fig:Proto}) are designed for efficiently estimating the per-group effective channels in each block, which depend on the intra-group training reflection vector. On the other hand, each group of IRS elements are further divided into subgroups with reducing size over the blocks, where  the intra-group training reflection vectors in the current block and all previous blocks (collectively called the \emph{intra-group training reflection matrix} which has an increasing dimension over the blocks and is identical for all the groups) are designed for efficiently resolving  the subgroup aggregated channels of each group, so that as the number of blocks becomes equal to that of IRS elements per group, all the subgroups constitute only one IRS element and thus all the IRS elements' individual channels can be resolved. 
\item In particular, we propose an efficient algorithm to design the basis training reflection matrix for each block, with any given intra-group training reflection matrix, for minimizing the mean-square error (MSE) of the per-group effective channel estimation, under the constraints of unit-modulus, discrete phase shifts, and full rank. Note that this problem is different from that considered in \cite{zheng2019intelligent} assuming the continuous phase shifts for IRS, since the constraint of discrete phase shifts renders this problem an  NP-hard problem, whose optimal solution can only be obtained by an exhaustive search. To reduce the computational complexity, we first show that the simple discrete Fourier transform (DFT)/Hadamard matrix is an optimal basis training reflection matrix in some special cases. Then, for other cases in general, we propose a low-complexity yet efficient method to sub-optimally solve this problem, called \emph{DFT-Hadamard-based} basis training reflection matrix design, which systematically constructs a \emph{near-orthogonal} basis  training reflection matrix based on either \emph{DFT-matrix quantization} or \emph{Hadamard-matrix truncation} depending on the phase-shifter resolution. 
\item Furthermore, for the intra-group channel estimation, we first propose a systematic approach to construct two types of subgroup partitions for dividing each group/subgroup of IRS elements into smaller-size subgroups over the blocks, called the \emph{symmetric} and \emph{asymmetric} subgroup partitions, respectively. Then we derive the conditions for the \emph{subgroup training reflection matrices} (which are determined by both the intra-group training reflection matrices and the subgroup partition scheme) to efficiently resolve the subgroup aggregated channels in each group over the  blocks, given any subgroup partition scheme.  Similar to the design for the basis training reflection matrix, the optimal subgroup training reflection matrices need to be found via the exhaustive search. To reduce the complexity, we propose a suboptimal method for designing the subgroup training reflection matrices over the blocks. The MSE of the resultant  progressive channel estimation in each block is derived in closed-form, which is shown to generally  depend on both the basis training and subgroup training reflection matrices. 
\item Last, based on the progressively refined IRS subgroup aggregated channels, we formulate an optimization problem to maximize the achievable rate  in each block by designing the corresponding per-subgroup based  passive beamforming for data transmission,  with the training overhead and the correlated channel estimation error taken into account. To reduce the complexity for finding the optimal solution via the exhaustive search, we propose a low-complexity \emph{successive refinement} algorithm with three properly-designed initializations, to obtain high-quality suboptimal solutions. Numerical results show that our proposed progressive channel estimation and passive beamforming designs can effectively  improve the achievable rate over the blocks. Moreover, the proposed designs significantly outperform the benchmark schemes under various practical setups.
\end{itemize}

The remainder of this paper is organized as follows. Section~\ref{Sec:Model} introduces the system model and the main ideas of the proposed progressive channel estimation and passive beamforming designs. The detailed designs for the per-group effective channel and intra-group channel estimations are presented in Sections~\ref{Sec:Inter-CE} and \ref{Sec:Intra-CE}, respectively. The algorithm for designing the progressive passive beamforming for data transmission  based on estimated channels is presented in Section~\ref{Sec:PBFOpt}. Numerical
results and discussions are provided in Section~\ref{Sec:Num}, followed by the conclusions given  in Section~\ref{Sec:Conc}.

\emph{Notations:} The superscripts $(\cdot)^{T}$, $(\cdot)^{H}$, $(\cdot)^{\dag}$, and $(\cdot)^{-1}$ denote respectively the operations of transpose, Hermitian transpose, element-wise conjugate, and matrix inversion. 
$\lceil \cdot\rceil$ and $\lfloor \cdot\rfloor$ denote respectively the ceiling and floor operations for a real number, and
$\angle(\cdot)$  denotes the phase of a complex number. Moreover, $\otimes$ and $\odot$  denote  the Kronecker and Hadamard products, respectively.{\footnote{\color{black}{For two matrices $\boldsymbol{A}\in\mathbb{C}^{n\times m}$ and $\boldsymbol{B}\in\mathbb{C}^{p\times q}$, their Kronecker product is defined as $$\boldsymbol{A}\otimes\boldsymbol{B}=\begin{bmatrix}
[\boldsymbol{A}]_{1,1}\boldsymbol{B} &   \cdots   & [\boldsymbol{A}]_{1,m}\boldsymbol{B}   \\
\vdots &    \ddots   & \vdots  \\
[\boldsymbol{A}]_{n,1}\boldsymbol{B} &   \cdots\  & [\boldsymbol{A}]_{n,m}\boldsymbol{B}  \\
\end{bmatrix}.$$ Moreover, under the conditions of $p=n$ and $q=m$, their Hadamard product is the entrywise product with each entry defined as $[\boldsymbol{A}\odot\boldsymbol{B}]_{i,j}=[\boldsymbol{A}]_{i,j} [\boldsymbol{B}]_{i,j}$.}}}  $|\cdot|$ denotes the cardinality for a set and the absolute value for a real number. For matrices, $\diag(\cdot)$ denotes a square diagonal matrix with the elements in $(\cdot)$ on its main diagonal, $[\cdot]_{i,j}$ denotes the ($i,j$)-th element of a matrix, $\lambda_{\max}({\cdot})$ denotes the maximum eigenvalue of a matrix, $\bm{I}_{M}$ denotes an identity matrix  with size $M\times M$, $\bm{1}_{N\times M}$ denotes an $N\times M$ all-one matrix, $\rank(\cdot)$ and $\tr(\cdot)$ represent the matrix rank and trace, respectively. The main symbols used in this paper are summarized in Table~I.


\begin{table*}[!htbp]
\centering
{\color{black}{
\caption{List of main symbols and their physical meanings.}
\begin{tabular}{| c | l ||c| l | }
\hline
\multirow{2}{*}{$N$}&\multirow{2}{*}{Total number of IRS elements}&\multirow{2}{*}{$M$}&\multirow{2}{*}{Number of IRS element groups}\\
&&&\\
\hline
\multirow{2}{*}{$L$}&\multirow{2}{*}{Number of IRS elements in each group}&\multirow{2}{*}{$I_0$}&\multirow{2}{*}{Number of time blocks}\\
&&&\\
\hline
\multirow{2}{*}{$i\in\mathcal{I}$}&\multirow{2}{*}{Time block index}& \multirow{2}{*}{$\tilde{m}\in\mathcal{M}$}&\multirow{2}{*}{Symbol index in each block}\\
&&&\\
\hline
\multirow{2}{*}{$m\in\mathcal{M}$}&\multirow{2}{*}{IRS element group  index}&\multirow{2}{*}{$k$}&\multirow{2}{*}{Subgroup index in each group}\\
&&&\\
\hline
\multirow{2}{*}{$\underline{\boldsymbol{h}}_{\rm UI}\in\mathbb{C}^{N\times1}$}&Element-wise baseband user-IRS channel &\multirow{2}{*}{$\underline{\boldsymbol{h}}_{\rm{IA}}^{H}\in\mathbb{C}^{1\times N}$}&\multirow{2}{*}{Element-wise baseband IRS-AP channel vector}\\
&  vector&&\\
\hline
\multirow{2}{*}{$\underline{\boldsymbol{h}}\in\mathbb{C}^{N\times 1}$}&Element-wise cascaded user-IRS-AP &\multirow{2}{*}{$\boldsymbol{\underline h}_m\in\mathbb{C}^{L\times 1}$}&\multirow{2}{*}{Element-wise channel vector of group $m$}\\
&channel vector&&\\
\hline
\multirow{2}{*}{$q\in\mathbb{C}$}&\multirow{2}{*}{Element-wise user-AP equivalent channel}&\multirow{2}{*}{$\boldsymbol{h}^{(i)}\in\mathbb{C}^{M\times 1}$}&Per-group
effective channel vector of $M$ groups\\
&&&in block $i$\\
\hline
 \multirow{2}{*}{ $h^{(i)}_m\in\mathbb{C}$}&\multirow{2}{*}{Effective channel of group $m$ in block $i$}&\multirow{2}{*}{\boldsymbol{${ \eta}}^{(i)}_m\in\mathbb{C}^{i\times 1}$}&Effective channel vector of group $m$ in the first \\
 & & &  $i$ blocks\\
\hline
\multirow{2}{*}{$\boldsymbol{g}^{(i)}_m\in\mathbb{C}^{i\times 1}$}&Subgroup aggregated channel vector of group&\multirow{2}{*}{$g^{(i)}_{m,k}\in\mathbb{C}$}&Aggregated channel of the elements  in the $k$-th\\
& $m$  in block $i$&& subgroup of group $m$\\
\hline
\multirow{2}{*}{ $\boldsymbol{{\hat g}}^{(i)}\in\mathbb{C}^{iM\times 1}$}&Subgroup aggregated channel vector of all &\multirow{2}{*}{$\mathbb{F}$}&\multirow{2}{*}{Reflection coefficient space for each element}\\
&groups  in block $i$&&\\
\hline
\multirow{2}{*}{$\underline{\boldsymbol{\Omega}}\in\mathbb{F}^{N\times N}$}&\multirow{2}{*}{Element-wise IRS  reflection matrix (diagonal)}&\multirow{2}{*}{$\underline{\boldsymbol{\theta}}^{H}\in\mathbb{F}^{1\times N}$}&\multirow{2}{*}{Element-wise IRS reflection vector}\\
&&&\\
\hline
 \multirow{2}{*}{$(\boldsymbol{\underline\theta}_{\rm t}^{(i)}[{\tilde m}])^{H}\in\mathbb{F}^{1\times N}$}  \multirow{2}{*}  &  Element-wise IRS training reflection
vector & \multirow{2}{*}{$\boldsymbol{\underline{\Theta}}^{(i)}_{\rm t}\in\mathbb{F}^{M\times N}$}&\multirow{2}{*}{Element-wise IRS training reflection matrix in block $i$}   
\tabularnewline \multirow{2}{*}  &  in the $\tilde{m}$-th symbol duration of block $i$ & &  \\
\hline
\multirow{2}{*}{$(\boldsymbol{\underline \theta}_{\rm{s}}[{\tilde m}])^{H}\in\mathbb{F}^{1\times N}$}&Element-wise basis training reflection vector&\multirow{2}{*}{$(\boldsymbol{\underline \theta}_{\rm{a}}^{(i)})^{H}\in\mathbb{F}^{1\times N}$}&Element-wise intra-group training reflection vector\\
&in the $\tilde{m}$-th symbol duration of each block&&in block $i$\\
\hline
 \multirow{2}{*}{$(\boldsymbol{\theta}_{{\rm{s}}}[{\tilde m}])^H\in\mathbb{F}^{1\times M}$}  \multirow{2}{*}  &  Group-wise basis training reflection vector & \multirow{2}{*}{$(\boldsymbol{\theta}^{(i)}_{{\rm{a}}})^H\in\mathbb{F}^{1\times L}$}& \multirow{2}{*}{Intra-group training reflection vector in block $i$}   \tabularnewline \multirow{2}{*}  & in the $\tilde{m}$-th symbol duration of each block & & \\
\hline
\multirow{2}{*}{$\boldsymbol{\Theta}_{\rm s}\in\mathbb{F}^{M\times M}$}&\multirow{2}{*}{Basis training reflection matrix in each block}&\multirow{2}{*}{$\boldsymbol{\Theta}^{(i)}_{{\rm{a}}}\in \mathbb{F}^{i\times L}$}&\multirow{2}{*}{Intra-group training reflection matrix in block $i$}\\
&&&\\
\hline
\multirow{2}{*}{$(\bm{\psi}_{\rm a}^{(i)})^{H}\in\mathbb{F}^{1\times i}$}&\multirow{2}{*}{Subgroup training reflection vector in block $i$}&\multirow{2}{*}{$\bm{\Psi}^{(i)}_{\rm a}\in\mathbb{F}^{i\times i}$}&\multirow{2}{*}{Subgroup training reflection matrix in block $i$}\\
&&&\\
\hline
\multirow{2}{*}{${\tilde {\boldsymbol \Psi}}_{\rm a}^{(i)}\in\mathbb{F}^{i\times i}$}& {Extended} subgroup training reflection matrix& \multirow{2}{*}{$(\boldsymbol{\phi}^{(i)})^{H}\in\mathbb{F}^{1\times iM}$}&\multirow{2}{*}{Passive beamforming vector in block $i$} \\
& in block $i$&&\\
\hline
\multirow{2}{*}{${\cal P}^{(i)}_m$}&\multirow{2}{*}{Subgroup partition for group $m$ in block $i$}&\multirow{2}{*}{${\cal S}^{(i)}_{m,k}$}&Element-index set  for 
 subgroup $k$ of group $m$\\
 &&&in block $i$\\
\hline
\end{tabular}}\label{Table}}
\end{table*}

\section{System Model}\label{Sec:Model}

For the purpose of exposition, we consider a basic IRS-aided single-user communication system as illustrated in Fig.~\ref{Fig:Syst}, where an IRS composed of a large number of $N$ passive reflecting elements, denoted by the set $\mathcal{N}=\{1, 2, \cdots, N\}$, is deployed in proximity  to a user for assisting its data transmission with an AP, both of which are equipped with a single antenna. The results in this paper can be readily extended to the more general system with multiple users served by the IRS (e.g., by applying orthogonal time/frequency division multiple access) and/or multiple antennas at the AP (by estimating their associated channels in parallel), which will be investigated in our future work.  {\color{black} The IRS is attached with a smart controller, which is implemented by e.g., field-programmable gate array (FPGA) powered by grid/battery  energy\cite{qingqing2019towards} and responsible for real-time adjustment of the amplitude and/or phase shift at each reflecting element as well as the low-rate information exchange between the IRS and AP within its connectivity range (e.g., hundreds of meters) via a separate reliable wireless link.}

\vspace{-5pt}
\subsection{Channel Model} 
For a typical low-mobility user  served by the IRS, we assume narrow-band quasi-static fading channels and focus on the uplink communication\footnote{The proposed designs in this paper  can be directly applied to the downlink communication by switching the roles of the user and AP, as well as the broadband communication over frequency-selective channels by employing OFDM (see, e.g., \cite{yang2019intelligent,zheng2019intelligent}).} in one transmission frame consisting of $I_0$ blocks, where all the channels remain constant within each frame. Note that $I_0$ is an arbitrary integer depending on the channel coherence time of the user. We further  assume that the direct link between the user and AP is blocked due to obstructions\footnote{If the direct link is non-negligible, the training symbols in the first block can be used to estimate the direct channel, without affecting  the main results in this paper.}, and denote $\underline{\boldsymbol{h}}_{\rm UI}\in\mathbb{C}^{N\times1}$ and $\underline{\boldsymbol{h}}_{\rm{IA}}^{H}\in\mathbb{C}^{1\times N}$ as the element-wise baseband (equivalent) channels of the user-IRS and IRS-AP links\footnote{In this paper, the underlined symbols (e.g., $\underline{\beta}_n$, $\underline{\boldsymbol{h}}_{\rm UI}$, and $\underline{\boldsymbol{\Omega}}$) refer to scalars/vectors/matrices with entries corresponding to individual IRS elements.}, respectively.
The reflection coefficients of all IRS elements can be represented by a diagonal matrix, denoted by $\underline{\boldsymbol{\Omega}}=\diag(\underline{\beta}_1 e^{j\underline{\omega}_1}, \underline{\beta}_2 e^{j\underline{\omega}_2}, \cdots, \underline{\beta}_N e^{j\underline{\omega}_N})$, where $\underline{\beta}_n\in[0,1]$ and $\underline{\omega}_n\in[0, 2\pi)$  denote respectively the reflection amplitude and phase shift at each element $n$. In practice, the phase shift of each element can only take a finite number of discrete values due to the hardware constraint \cite{qingqing2019towards}. Specifically, let $b$ denote the number of controlling bits per element  and $K=2^{b}$ denote the number of discrete phase-shift levels. By uniformly quantizing  the continuous phase shift in the range of $[0, 2\pi)$, the set of all possible discrete phase shifts for each element can be represented by  $\mathcal{F}\triangleq\{0, \Delta\omega, \cdots, (K-1)\Delta\omega\}$,
where $\Delta\omega=2\pi/K$. To ease  the design of reflection coefficients and maximize  the reflected signal power, we consider the \emph{full-reflection} at the IRS for both channel training and data transmission, where the reflection amplitude at each element is set to be its maximum, i.e., $\underline{\beta}_n=1, \forall n\in\mathcal{N}$. Then the reflection coefficient for each element can be represented by the set  $\mathbb{F}=\{e^{j\underline{\omega}}| \underline{\omega}\in\mathcal{F}\}$.
Similar to \cite{wu2019intelligent}, the equivalent channel from the user to the AP depends on $\underline{\boldsymbol{\Omega}}$ and can be expressed as
\begin{equation}
q(\underline{\boldsymbol{\Omega}})= \underline{\boldsymbol{h}}_{\rm{IA}}^{H}~ \underline{\boldsymbol{\Omega}} ~\underline{\boldsymbol{h}}_{\rm{UI}}.\label{Eq:Chan}
\end{equation}
Let $\underline{\boldsymbol{h}}\triangleq\diag(\underline{\boldsymbol{h}}_{\rm{IA}}^{H}) \underline{\boldsymbol{h}}_{\rm{UI}}\in\mathbb{C}^{N\times 1}$ denote the element-wise cascaded user-IRS-AP channel vector without the phase-shift adjustment, and $\underline{\boldsymbol{\theta}}^{H}\triangleq \l[e^{j\underline{\omega}_1}, \cdots, e^{j\underline{\omega}_N}\r]\in\mathbb{F}^{1\times N}$ denote the element-wise IRS reflection vector with $\underline{\theta}^{\dag}_n\triangleq e^{j\underline{\omega}_n}$ being the reflection coefficient of element $n$, $n\in \mathcal{N}$. Note that the product of any two elements in $\underline{\theta}^{\dag}_n$ is another element in it due to the phase periodicity over $\pm 2\pi$. Then the equivalent channel given in \eqref{Eq:Chan} can be rewritten as $q(\boldsymbol{{\underline\theta}})= \boldsymbol{{\underline\theta}}^{H}\boldsymbol{\underline h}$. It is worth noting that if the perfect CSI of $\boldsymbol{\underline h}$ is available, the optimal design of IRS  passive beamforming with discrete phase shifts  for rate maximization can be obtained by using the techniques in \cite{wu2019beamforming}. However, such perfect CSI is practically difficult to obtain for IRS  as explained in Section~\ref{Sec:Intro}. Thus, we propose a new approach to practically design the IRS channel estimation jointly with  passive beamforming for data transmission  based on the estimated channels, both under the constraint of discrete phase shifts, as will be detailed in the next subsection.

\vspace{-5pt}
\subsection{Proposed Progressive Channel Estimation and Passive Beamforming Design}
We consider a practical  protocol for the IRS-aided uplink communication as illustrated in Fig.~\ref{Fig:Proto}, where each transmission frame consists of $I_0$ consecutive blocks and each block consists of $M_0$ symbols that are divided into two portions for executing the channel training with the first $M$ symbols and the data transmission with the remaining $M_0-M$ symbols, respectively. Based on this protocol, we propose a novel \emph{hierarchical} training reflection design to \emph{progressively} refine the channel estimation for IRS elements over the blocks.
The estimated channels in each block  are then used for designing the corresponding passive beamforming  for data transmission to improve the achievable rate for the user over the blocks. 
\subsubsection{Channel Training}
Consider each block $i\in \mathcal{I}\triangleq\{1, 2,\cdots, I_0\}$. During the channel training in this block, the user consecutively sends $M$ pilot symbols to the AP, where  the  IRS  reflection coefficients are properly set to assist the channel estimation at the AP. {\color{black}Let  $x_{\rm t}[{\tilde m}]\in \mathbb{C}$ denote the transmitted training symbol (common for all blocks of $i$), $(\boldsymbol{\underline\theta}_{\rm t}^{(i)}[{\tilde m}])^{H}\in\mathbb{F}^{1\times N}$ denote the element-wise IRS training reflection vector in the $\tilde{m}$-th symbol duration of block $i$ with $\tilde{m}\in\mathcal{M}\triangleq\{1, 2, \cdots, M\}$, and $z^{(i)}_{\rm t}[{\tilde m}]$ denote the additive white Gaussian noise  at the receiver with zero mean and variance $\sigma^2$. 
By stacking $M$ consecutive received signals during the channel training in block $i$, i.e., $\boldsymbol{y}^{(i)}_{\rm t}\triangleq[y^{(i)}_{\rm t}[1],y^{(i)}_{\rm t}[2], \cdots, y^{(i)}_{\rm t}[M]]^{T}$, and defining $\boldsymbol{\underline{\Theta}}^{(i)}_{\rm t}\triangleq[\boldsymbol{\underline{\theta}}_{\rm t}^{(i)}[1], \boldsymbol{\underline{\theta}}_{\rm t}^{(i)}[2], \cdots, \boldsymbol{\underline{\theta}}_{\rm t}^{(i)}[M]]^H$, the received signal vector  can be compactly written as
\vspace{-5pt}
\begin{equation}
\boldsymbol{y}^{(i)}_{\rm t}= \boldsymbol{X}_{\rm t} \boldsymbol{q}^{(i)}_{\rm t}  +\boldsymbol{z}^{(i)}_{\rm t},
\label{Eq:TrainRecComp}
\end{equation}
where $\boldsymbol{X}_{\rm t}\triangleq\diag\l(x_{\rm t}[1], x_{\rm t}[2], \cdots, x_{\rm t}[M]\r)$, $\boldsymbol{q}^{(i)}_{\rm t}\triangleq[q^{(i)}_{\rm t}[1], q^{(i)}_{\rm t}[2], \cdots, q^{(i)}_{\rm t}[M]]^{T}=\boldsymbol{\underline{\Theta}}^{(i)}_{\rm t}\boldsymbol{\underline h}$ with $q^{(i)}_{\rm t}[{\tilde m}]\triangleq(\boldsymbol{\underline{\theta}}_{\rm t}^{(i)}[{\tilde m}])^{H} \boldsymbol{\underline h}$, and  $\boldsymbol{z}^{(i)}_{\rm t}\triangleq[z^{(i)}_{\rm t}[1], z^{(i)}_{\rm t}[2], \cdots, z^{(i)}_{\rm t}[M]]^{T}$.}


Based on  \eqref{Eq:TrainRecComp},   the least-square (LS) estimation of the equivalent channel vector in each block $i$, denoted by $\boldsymbol{{\hat q}}^{(i)}_{\rm t}$, can be obtained as $\boldsymbol{{\hat q}}^{(i)}_{\rm t}=\boldsymbol{X}_{\rm t}^{-1}\boldsymbol{y}^{(i)}_{\rm t}$.
Thus, if $M=N$, the cascaded user-IRS-AP channels, $\boldsymbol{\underline h}$, can be completely resolved as $\boldsymbol{\underline {\hat h}}=(\boldsymbol{\underline{\Theta}}^{(i)}_{\rm t})^{-1}\boldsymbol{{\hat q}}^{(i)}_{\rm t}$ (i.e., the all-at-once channel estimation). However, in practice, we have $M\ll N$ and thus we propose to divide the $N$ IRS elements into $M$ groups, each consisting of $L\triangleq N/M$ adjacent elements (assumed to be an integer for convenience)  by exploiting the potential channel correlation among them \cite{yang2019intelligent}. As such, in each block $i$, we can estimate $M$  per-group effective channels (to be specified later), denoted by $\boldsymbol{h}^{(i)}\triangleq[h^{(i)}_1, h^{(i)}_2, \cdots, h^{(i)}_M]^{T}\in\mathbb{C}^{M\times 1}$,  where $h^{(i)}_m$ with $m\in\mathcal{M}$ denotes the effective channel of group $m$ in block $i$. 

\begin{figure*}[t]
\begin{center}
\includegraphics[height=10cm]{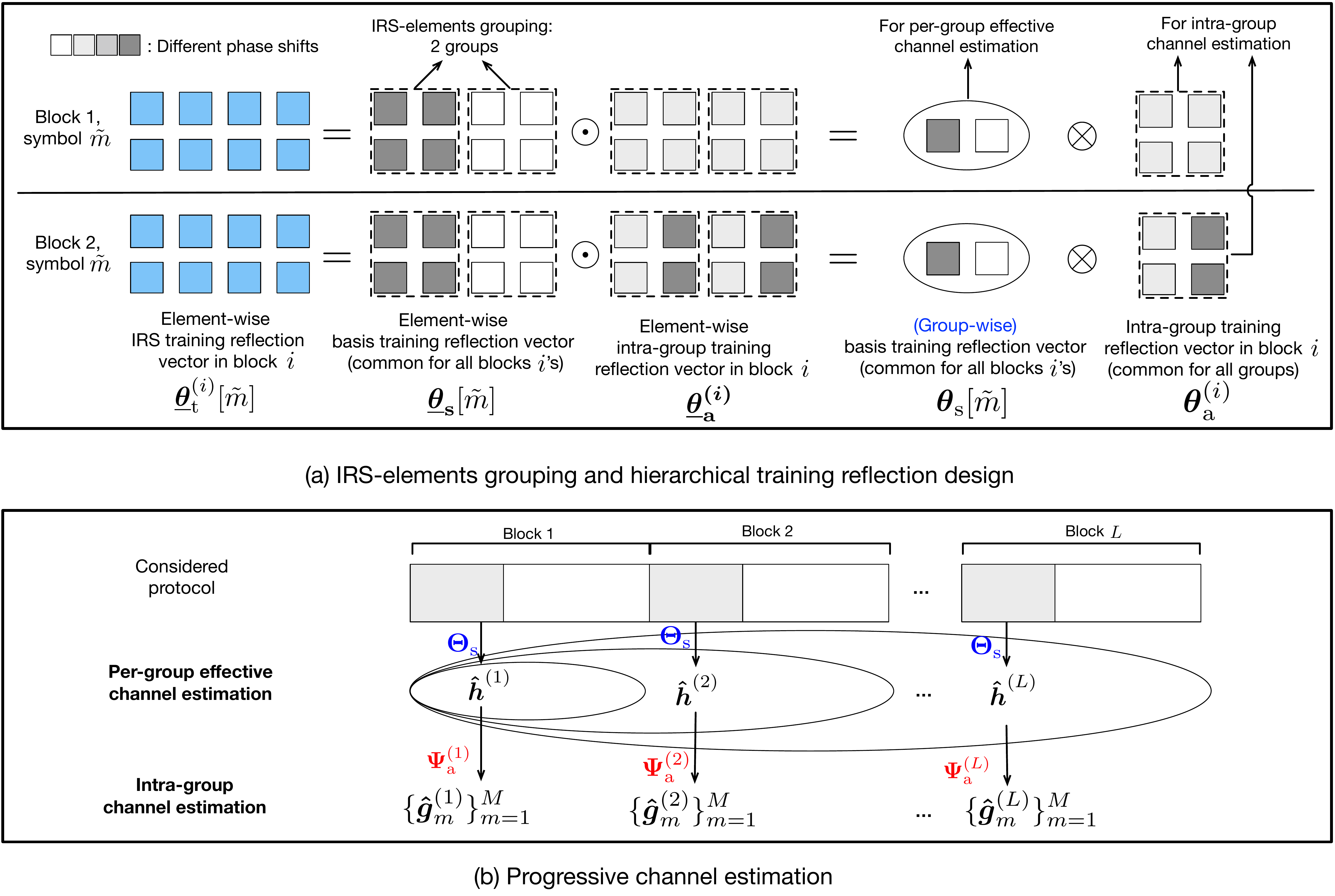}
\caption{Proposed progressive channel estimation by hierarchical training reflection design.}
\label{Fig:GroupHier}
\end{center}
\end{figure*}

Since each transmission frame consists of more than one block, we further propose a novel hierarchical training reflection design to progressively estimate the cascaded IRS channels over the blocks. Specifically,  we first write the element-wise IRS training reflection vector, $(\boldsymbol{\underline{\theta}}_{\rm t}^{(i)}[{\tilde m}])^{H}$, as a Hadamard product of  two vectors, i.e., $(\boldsymbol{\underline \theta}^{(i)}_{{\rm t}}[{\tilde m}])^{H}=  (\boldsymbol{\underline \theta}_{\rm{s}}[{\tilde m}])^{H}\odot (\boldsymbol{\underline \theta}_{\rm{a}}^{(i)})^{H}$, as illustrated in Fig.~\ref{Fig:GroupHier}(a).  {\color{black}Therein, $(\boldsymbol{\underline \theta}_{\rm{s}}[{\tilde m}])^{H}\in\mathbb{F}^{1\times N}$ is called the \emph{element-wise basis training reflection} vector, which is assumed to be identical for all the blocks in each frame such that the proposed per-group effective channel estimation (to be specified later) applies to all blocks.} In each block, the element-wise basis training reflection coefficients for all elements in each group are set identical for each training symbol  but can vary over different training symbols for estimating the per-group effective channels.  In contrast, $(\boldsymbol{\underline \theta}_{\rm{a}}^{(i)})^{H}$ is named the \emph{element-wise intra-group training reflection} vector, which is set identical for all training symbols in each block $i$, but can vary over different blocks of $i$. {\color{black}In addition, the element-wise intra-group training reflection coefficients are set to be identical for all groups in each block such that the proposed intra-group channel estimation method in each block (to be specified  later) applies to all groups.} The aim of designing $(\boldsymbol{\underline \theta}_{\rm{a}}^{(i)})^{H}$ is that upon each block $i$, the intra-group training reflection vectors, $\{(\boldsymbol{\underline \theta}_{\rm{a}}^{(j)})^{H}\}_{j=1}^{i}$, applied from blocks $1$ to $i$ can help resolve more subgroup  channels in each group as $i$ increases. Note that in block $2$, each group of elements are divided into $2$ subgroups, then in block $3$, one of the two subgroups is further divided into $2$ subgroups, and so on; as a result, there are $i$ subgroups in each group in block $i$, for $i>1$. Moreover, for all elements in each group/subgroup, we apply the same reflection coefficients so that their effective/aggregated  channels can be resolved in each block $i$. For example, consider a simple case where the IRS has $N=4$ elements and each block consists of $M=2$ training symbols. Since $M=2$,  the IRS elements are divided into $2$ groups of equal size, which are denoted by the element-index sets, e.g.,  $\{1, 2\}$ and $\{3,4\}$, respectively. Then in block $2$, the AP is able to resolve two subgroup channels  from each group, i.e., $\{1\}$ and $\{2\}$ from group $1$ and $\{3\}$ and $\{4\}$ from group $2$; thus,  all the four IRS elements' channels are resolved.
Based on the above, the element-wise IRS training reflection vector during the ${\tilde m}$-th symbol of each block $i$ can be expressed as below and illustrated in Fig.~\ref{Fig:GroupHier}(a), which is the Kronecker product of a \emph{(group-wise) basis training reflection vector}, denoted by $(\boldsymbol{\theta}_{{\rm{s}}}[{\tilde m}])^H$, and an \emph{intra-group training reflection vector} in the $i$-th block (common for all groups), denoted by $(\boldsymbol{\theta}^{(i)}_{{\rm{a}}})^H$.
\vspace{-5pt}
\begin{align}
\!\!\!(\boldsymbol{\underline \theta}^{(i)}_{{\rm t}}[{\tilde m}])^{H}&=  (\boldsymbol{\underline \theta}_{\rm{s}}[{\tilde m}])^{H}\odot (\boldsymbol{\underline \theta}_{\rm{a}}^{(i)})^{H}\nn\\
&\triangleq \l[\theta_{{\rm{s}}, 1}[{\tilde m}]\otimes \boldsymbol{1}_{1\times L}, ~\theta_{{\rm{s}}, 2}[{\tilde m}]\otimes \boldsymbol{1}_{1\times L},\cdots, \right.\nn\\ 
&\qquad\quad \left.\quad\theta_{{\rm{s}}, M}[{\tilde m}] \otimes \boldsymbol{1}_{1\times L} \r]^{\dag}\odot
\l[(\boldsymbol{\theta}^{(i)}_{{\rm{a}}})^H \otimes \boldsymbol{1}_{1\times M}\r]\nn\\
&=\l[\underbrace{(\theta_{{\rm{s}}, 1}[{\tilde m}])^{\dag} (\boldsymbol{\theta}^{(i)}_{{\rm{a}}})^H}_{{\rm group}~1},  \underbrace{(\theta_{{\rm{s}}, 2}[{\tilde m}])^{\dag} (\boldsymbol{\theta}^{(i)}_{{\rm{a}}})^H}_{{\rm group}~2}, \cdots, \right.\nn\\ 
&\qquad\quad \left.\qquad \qquad\qquad\qquad\underbrace{(\theta_{{\rm{s}}, M}[{\tilde m}])^{\dag} (\boldsymbol{\theta}^{(i)}_{{\rm{a}}})^H}_{{\rm group}~M}\r]\label{Eq:intra-pergroupDecomp}\nn\\
&= (\boldsymbol{\theta}_{{\rm{s}}}[{\tilde m}])^H\otimes(\boldsymbol{\theta}^{(i)}_{{\rm{a}}})^H, \forall \tilde{m}\in\mathcal{M}, i\in\mathcal{I},
\end{align}
where $(\boldsymbol{\theta}_{{\rm{s}}}[{\tilde m}])^H\triangleq [ \theta_{{\rm{s}}, 1}[{\tilde m}], \theta_{{\rm{s}}, 2}[{\tilde m}], \cdots, \theta_{{\rm{s}}, M}[{\tilde m}]]^{\dag}\in\mathbb{F}^{1\times M}$  and $(\boldsymbol{\theta}^{(i)}_{{\rm{a}}})^H\triangleq [ \theta^{(i)}_{{\rm{a}}, 1}, \theta^{(i)}_{{\rm{a}}, 2}, \cdots, \theta^{(i)}_{{\rm{a}}, L}]^{\dag}\in\mathbb{F}^{1\times L}$. 
In particular, it can be observed from \eqref{Eq:intra-pergroupDecomp} that for each symbol $\tilde{m}$ in each block $i$, the training reflection vector of each group is the superposition of a common basis training reflection coefficient for this group for symbol $\tilde{m}$ to the intra-group training reflection vector in block $i$. By partitioning the IRS elements' channels as $\boldsymbol{\underline h}=[\boldsymbol{\underline h}_1^{H}, \boldsymbol{\underline h}_2^{H}, \cdots, \boldsymbol{\underline h}_M^{H}]^{H}$, where $\boldsymbol{\underline h}_m\in\mathbb{C}^{L\times 1}$ denotes the element-wise channels of group $m$, then the equivalent channel during the $\tilde{m}$-th symbol of block $i$ can be rewritten as
\begin{align}
q^{(i)}_{\rm t}[{\tilde m}]
& =\l[(\theta_{{\rm{s}}, 1}[{\tilde m}])^{\dag} (\boldsymbol{\theta}^{(i)}_{{\rm{a}}})^H, \cdots, (\theta_{{\rm{s}}, M}[{\tilde m}])^{\dag} (\boldsymbol{\theta}^{(i)}_{{\rm{a}}})^H\r]\nn \\
&\qquad\qquad\qquad\qquad\qquad\qquad\times [\boldsymbol{\underline h}_1^{H},\cdots, \boldsymbol{\underline h}_M^{H}]^{H}\nn\\
&=\sum_{m=1}^{M}\l((\theta_{{\rm{s}}, m}[{\tilde m}])^{\dag} (\boldsymbol{\theta}^{(i)}_{{\rm{a}}})^H \boldsymbol{\underline h}_{m}\r)\nn\\
&\triangleq(\boldsymbol{\theta}_{{\rm{s}}}[{\tilde m}])^H \boldsymbol{h}^{(i)},\label{Eq:GroupC}
\end{align}
where each per-group effective channel $h^{(i)}_m$ in $\boldsymbol{h}^{(i)}$ depends on the intra-group training reflection vector of block $i$, i.e., 
\vspace{-5pt}
\begin{equation}\label{Eq:effePerChan}
h^{(i)}_m=(\boldsymbol{\theta}^{(i)}_{{\rm{a}}})^H \boldsymbol{\underline h}_m,
\end{equation}
As such, $\boldsymbol{y}^{(i)}_{\rm t}$ in \eqref{Eq:TrainRecComp} can be re-expressed as
\begin{align}\label{Eq:IntraReciSig}
\boldsymbol{y}^{(i)}_{\rm t}=\boldsymbol{X}_{\rm t} \boldsymbol{\Theta}_{\rm s} \boldsymbol{h}^{(i)}+\boldsymbol{z}^{(i)}_{\rm t},
\end{align} 
where $\boldsymbol{\Theta}_{\rm s}\triangleq\l[\boldsymbol{\theta}_{{\rm{s}}}[1], \boldsymbol{\theta}_{{\rm{s}}}[2], \cdots, \boldsymbol{\theta}_{{\rm{s}}}[{\tilde m}]\r]^H\in\mathbb{F}^{M\times M}$ is collectively defined as the \emph{basis training reflection matrix}. In the following, we summarize the procedures of the proposed progressive channel estimation in each block $i$, as illustrated in Fig.~\ref{Fig:GroupHier}(b), by considering the case of $I_0=L$ (i.e., $\mathcal{I}=\{1,2, \cdots, L\}$) in the sequel of this paper for convenience, such that all IRS elements' individual channels can be resolved by block $I_0=L$, since $ML=N$.\footnote{
It is worth noting that  our proposed design can be easily applied  to the case of $I_0< L$, by simply adopting the design for the first $I_0$ out of the total $L$ blocks; while for the case of  $I_0>L$, since all the element-wise channels can be estimated by block $L$, we can assume for simplicity that the training symbols in the remaining $I_0-L$ blocks are unused or used for data transmission.}\\
\textbf{\underline{Per-group effective channel estimation}}: 
According to \eqref{Eq:IntraReciSig}, the AP estimates $M$ per-group effective channels in block $i$, $\boldsymbol{h}^{(i)}$,  from the received signal vector,  $\boldsymbol{y}^{(i)}_{\rm t}$, with  the given basis training reflection matrix $\boldsymbol{\Theta}_{\rm s}$, i.e., 
\begin{align}
&(\text{Per-group effective channel estimation})\nn\\
&\boldsymbol{y}^{(i)}_{\rm t}\overset{\boldsymbol{\Theta}_{\rm s}}{\Longrightarrow} \boldsymbol{{\hat h}}^{(i)}, \quad \forall i\in\mathcal{I},
\end{align}
where $\boldsymbol{{\hat h}}^{(i)}$ denotes the estimated per-group effective channels in block $i$. The details will be given in Section~\ref{Sec:Inter-CE}. \\
\textbf{\underline{Intra-group channel estimation}}: 
Based on the estimated per-group effective channels from blocks $1$ to $i$ (i.e., $\{\boldsymbol{{\hat h}}^{(j)}\}_{j=1}^{i}$), the AP can estimate $iM$ effective channels (with $i$ effective channels per group) with the designed intra-group training reflection matrices. As the intra-group channel estimation design applies to all groups, we  consider a typical group $m$ for ease of elaboration. It can be observed from \eqref{Eq:effePerChan} that the effective channel of group $m$ in each block $i$ is a linear combination of its element-wise channel vector  and the  intra-group training reflection vector in this block. This key observation indicates that by properly designing the intra-group training reflection vectors over blocks, we can progressively resolve the IRS elements' channels in group $m$.   Let $\boldsymbol{{ \eta}}^{(i)}_m=[{h}^{(1)}_m, \cdots, {h}^{(i)}_m]^{T}\in\mathbb{C}^{i\times 1}$ denote the \emph{stacked} effective channels of group $m$ in the first $i$ blocks, which can be expressed as follows according to \eqref{Eq:effePerChan}:
\begin{align}\label{Eq:IntraCERela}
\boldsymbol{{ \eta}}^{(i)}_m =[\boldsymbol{\theta}^{(1)}_{\rm a}, \cdots, \boldsymbol{\theta}^{(i)}_{\rm a}]^H
\boldsymbol{\underline h}_m  \triangleq \boldsymbol{\Theta}^{(i)}_{{\rm{a}}}\boldsymbol{\underline h}_m,
\end{align}
where $\boldsymbol{\Theta}^{(i)}_{{\rm{a}}}\in \mathbb{F}^{i\times L}$ is named the \emph{intra-group training reflection matrix} for block $i$, evolving as 
\begin{align}\label{Eq:Coupling}
\boldsymbol{\Theta}^{(1)}_{{\rm{a}}}=({\boldsymbol{\theta}^{(1)}_{\rm a}})^{H},~~~\boldsymbol{\Theta}^{(i)}_{{\rm{a}}}=\begin{bmatrix}
\boldsymbol{\Theta}^{(i-1)}_{{\rm{a}}} \\
({\boldsymbol{\theta}^{(i)}_{\rm a}})^{H}
\end{bmatrix}, 1< i \le L.
\end{align}

Note that in block $i$,  we can at most resolve $i$ subgroup aggregated channels from $\boldsymbol{{ \eta}}^{(i)}_m$ for  group $m$.   Let $\boldsymbol{g}^{(i)}_m\triangleq[g^{(i)}_{m,1}, g^{(i)}_{m,2}, \cdots, g^{(i)}_{m,i}]^{T}\in\mathbb{C}^{i\times 1}$ denote the  subgroup aggregated channels of group $m$ in block $i$ (including the group aggregated channel for the case of $i=1$ as well), where $g^{(i)}_{m,k}$ with $k\le i$ denotes the aggregated channel of the elements in the $k$-th subgroup (i.e., the sum of the elements' individual channels in each subgroup) of group $m$. 
 Moreover,  to resolve the subgroup aggregated channels, we define $(\bm{\psi}_{\rm a}^{(i)})^{H}\triangleq[\psi^{(i)}_{{\rm a},1}, \psi^{(i)}_{{\rm a},2}, \cdots, \psi^{(i)}_{{\rm a}, i}]^{\dag}\in\mathbb{F}^{1\times i}$ as the \emph{subgroup training reflection vector} for the $i$ subgroup aggregated channels to be resolved in block $i$, which is common for all groups, and $(\psi^{(i)}_k)^{\dag}$ with $k\leq i$ represents the (common) reflection coefficient for all elements in the $k$-th subgroup applied in block $i$ (see Example 1 below). Note that each $({\boldsymbol{\theta}^{(i)}_{\rm a}})^{H}$ specifies $(\bm{\psi}_{\rm a}^{(i)})^{H}$, if the subgroup partition is given for block $i$ (to be specified in Section~\ref{Sec:Intra-CE}).   Furthermore, we define $\bm{\Psi}^{(i)}_{\rm a}\in\mathbb{F}^{i\times i}$ as the \emph{subgroup training reflection matrix} for block $i$, designed for resolving the subgroup aggregated channels $\boldsymbol{g}^{(i)}_m$ from $\boldsymbol{{ \eta}}^{(i)}_m$. An illustrative example is provided as follows to demonstrate the construction of $\bm{\Psi}^{(i)}_{\rm a}$, given a subgroup partition that determines the subgroup training reflection vector $(\bm{\psi}_{\rm a}^{(i)})^{H}$ and the corresponding subgroup aggregated channels $\boldsymbol{g}^{(i)}_m$'s; while the details for the subgroup partition and the design of $\bm{\Psi}^{(i)}_{\rm a}$ will be given in Section~\ref{Sec:Intra-CE}. Note that $\bm{\Psi}^{(i)}_{\rm a}$ (as well as $\boldsymbol{\Theta}^{(i)}_{{\rm{a}}}$) can be designed off-line and stored at the IRS for real-time channel training. 
\vspace{-5pt}
\begin{example}[Construction of the subgroup training reflection matrix]\label{Exp:SubTrain}\emph{Without loss of generality, we consider the intra-group channel estimation for group $m=1$ with $L$ elements. In block $1$, all elements in this group share the same subgroup training reflection coefficient $(\psi^{(1)}_{{\rm a},1})^{\dag}$. Then the  effective channel of group $m$ in this block is given by $h^{(1)}_{m}=(\psi^{(1)}_{{\rm a},1})^{\dag}g^{(1)}_{m,1}\triangleq \bm{\Psi}^{(1)}_{\rm a} g^{(1)}_{m,1}$, where $g^{(1)}_{m,1}=\sum_{n=1}^{L} \underline{h}_{n}$. In block $2$, we assume that the $L$ elements are partitioned into two subgroups consisting of $v$ and $L-v$ elements, respectively, where $0<v<L$. As such, the effective channel of group $m$ in block $2$ is given by 
\begin{equation}\label{Eq:block2}
h^{(2)}_{m}=(\psi^{(2)}_{{\rm a},1})^{\dag}g^{(2)}_{m,1}+(\psi^{(2)}_{{\rm a},2})^{\dag}g^{(2)}_{m,2},
\end{equation} where $g^{(2)}_{m,1}=\sum_{n=1}^{v} \underline{h}_{n}$ and $g^{(2)}_{m,2}=\sum_{n=v+1}^{L} \underline{h}_{n}$. Given $g^{(2)}_{m,1}$ and $g^{(2)}_{m,2}$, we can rewrite the effective channel of group $m$ in block $1$ as 
\vspace{-5pt}
\begin{equation}\label{Eq:block1}
h^{(1)}_{m}=(\psi^{(1)}_{{\rm a},1})^{\dag} g^{(2)}_{m,1} +(\psi^{(1)}_{{\rm a},1})^{\dag}g^{(2)}_{m,2},
\end{equation} 
since $g^{(1)}_{m,1}=g^{(2)}_{m,1}+g^{(2)}_{m,2}$. Thus,  the stacked effective channels of group $m$ in the first $2$ blocks, $\boldsymbol{{ \eta}}^{(2)}_m$ as given in \eqref{Eq:IntraCERela}, can be equivalently expressed as follows by combing \eqref{Eq:block2} and \eqref{Eq:block1}:
\vspace{-5pt}
\begin{equation}\label{Eq:GeneSubMatrix}
\boldsymbol{{ \eta}}^{(2)}_m = \begin{bmatrix}
\psi^{(1)}_{{\rm a},1}  & \psi^{(1)}_{{\rm a},1} \\
\psi^{(2)}_{{\rm a},1}  & \psi^{(2)}_{{\rm a},2}  \\
\end{bmatrix}^{\dag}
\begin{bmatrix}
g^{(2)}_{m,1}   \\
g^{(2)}_{m,2}   \\
\end{bmatrix}\triangleq\bm{\Psi}^{(2)}_{\rm a}\boldsymbol{g}^{(2)}_m.
\end{equation}
Following the similar procedures as for constructing $\bm{\Psi}^{(2)}_{\rm a}$, we can obtain $\bm{\Psi}^{(i)}_{\rm a}$ for $2<i\le L$.}
\end{example}
Accordingly,  for each group $m$, the intra-group channel estimation in block $i$ can resolve the subgroup aggregated channels $\boldsymbol{g}^{(i)}_m$  from the stacked estimated effective channels in the first $i$ blocks, $\boldsymbol{{\hat \eta}}^{(i)}_{m}$, by properly designing the subgroup training reflection matrix  $\boldsymbol{\Psi}^{(i)}_{{\rm{a}}}$, i.e.,
\begin{align}\label{Eq:IntraCEDef}
&(\text{Intra-group channel estimation})\nn\\
&\boldsymbol{{\hat \eta}}^{(i)}_{m}\overset{\boldsymbol{\Psi}^{(i)}_{{\rm{a}}}}{\Longrightarrow} \boldsymbol{{\hat g}}^{(i)}_m, ~~~\forall m\in\mathcal{M}, ~~\forall i\in\mathcal{I},
\end{align}
where $\boldsymbol{{\hat g}}^{(i)}_m$ denotes  the estimated subgroup aggregated channels for group $m$.
The estimated subgroup aggregated channels of all groups in block $i$ are concatenated as  $\boldsymbol{{\hat g}}^{(i)}=[(\boldsymbol{{\hat g}}^{(i)}_1)^{H}, \cdots, (\boldsymbol{{\hat g}}^{(i)}_M)^{H}]^{H}\in\mathbb{C}^{iM\times 1}$. 

\subsubsection{Passive Beamforming}
In each block $i$, given the estimated subgroup aggregated channels of all groups, $\boldsymbol{{\hat g}}^{(i)}$, the AP optimizes the  passive beamforming for data transmission and then sends the corresponding phase-shift values  to the IRS controller for implementation\footnote{For simplicity, we assume that such feedback is error-free and has zero delay, while the proposed design applies to imperfect feedback in practice as well. For example, suppose the feedback incurs one block delay, then the proposed design can be simply modified such that in each block i, the IRS implements the passive beamforming designed by the AP in block $i-1$, for $i=2,\cdots,L$.}.  Let  $(\boldsymbol{\phi}^{(i)})^{H}\in\mathbb{F}^{1\times iM}$ denote the passive beamforming vector  in block $i$. Note that since all elements in each subgroup apply the same reflection coefficient (similar to the channel training case) for data transmission, the size of $(\boldsymbol{\phi}^{(i)})^{H}$  needs to be equal to the total number of subgroups with resolved aggregated channels, i.e., $iM$. Moreover, we define  $\boldsymbol{{ g}}^{(i)}_{\rm e}\triangleq\boldsymbol{{ g}}^{(i)}-\boldsymbol{{\hat g}}^{(i)}$ as the channel estimation error in $\boldsymbol{{\hat g}}^{(i)}$. Then the received data signal at the AP in block $i$ can be written as
\vspace{-5pt}
\begin{align}
y^{(i)} &=  (\boldsymbol{\phi}^{(i)})^{H} \boldsymbol{g}^{(i)} x^{(i)} +z^{(i)}\nn \\
&= (\boldsymbol{\phi}^{(i)})^{H} \l(\boldsymbol{{\hat g}}^{(i)}-\boldsymbol{{ g}}^{(i)}_{\rm e}\r)x^{(i)}  +z^{(i)}\nn\\
&= (\boldsymbol{\phi}^{(i)})^{H} \boldsymbol{{\hat g}}^{(i)}x^{(i)}- (\boldsymbol{\phi}^{(i)})^{H}\boldsymbol{{ g}}^{(i)}_{\rm e} x^{(i)}  +z^{(i)},
\end{align}
where  $x^{(i)}$ is the transmitted signal with zero mean and power  $P$, and $(\boldsymbol{\phi}^{(i)})^{H}\boldsymbol{{ g}}^{(i)}_{\rm e} x^{(i)}$ is the additional  interference due to the channel estimation error, whose power depends on the passive beamforming vector  $(\boldsymbol{\phi}^{(i)})^{H}$ as well as the channel estimation error $\boldsymbol{{ g}}^{(i)}_{\rm e}$. As such, the average achievable rate of each block $i$ in bits per second per Hertz (bps/Hz) is given by \cite{samardzija2003pilot}
\begin{align}
R^{(i)}&=\frac{M_0-M}{M_0}\times\nn\\
&\log_2\l(1+\frac{P \l|(\boldsymbol{\phi}^{(i)})^{H}\boldsymbol{{\hat g}}^{(i)}\r|^2}{\Gamma \l(P \mathbb{E}\l[\l|(\boldsymbol{\phi}^{(i)})^{H}\boldsymbol{{g}}_{\rm e}^{(i)}\r|^2\r]+\sigma^2\r)}\r),\forall i\in\mathcal{I},\label{Eq:Rate}
\end{align} 
where $\Gamma\ge 1$ stands for the achievable rate gap due to a practical
modulation and coding scheme. Note that the achievable rate, $R^{(i)}$ in \eqref{Eq:Rate}, is determined by 
the following signal-to-interference-plus-noise ratio (SINR):
\vspace{-5pt}
\begin{align}
\gamma(\boldsymbol{\phi}^{(i)})&=\frac{P \l|(\boldsymbol{\phi}^{(i)})^{H}\boldsymbol{{\hat g}}^{(i)}\r|^2}{P \mathbb{E}\l[\l|(\boldsymbol{\phi}^{(i)})^{H}\boldsymbol{{ g}}_{\rm e}^{(i)}\r|^2\r]+\sigma^2},~~~ \forall i\in\mathcal{I}.\label{Eq:SINR}
\end{align} 
Our objective is to optimize the passive beamforming with discrete phase shifts at the IRS to maximize the achievable rate for data transmission in each block $i$ (see Section~\ref{Sec:PBFOpt} for the details), so that as $i$ increases, the achievable rate will be progressively improved as more IRS subgroup aggregated channels are resolved.

Last, we summarize in Algorithm~\ref{Alg:Prog} the main procedures of the proposed progressive channel estimation and passive beamforming designs.

\vspace{-8pt}
\section{Per-Group Effective Channel Estimation}\label{Sec:Inter-CE}
In this section, we consider the per-group effective channel estimation in each block with any given intra-group training reflection design. An optimization problem is formulated and solved to minimize the MSE of the LS channel estimation by designing the IRS basis training reflection matrix.  
\vspace{-8pt}
\subsection{Problem Formulation}
As the per-group effective channel estimation design applies to all blocks, we drop the block index (i.e., the superscript $(i)$) in this section for notational brevity. First, we can observe from \eqref{Eq:IntraReciSig} that,
if  $\boldsymbol{\Theta}_{\rm s}$ is of full-rank, the  LS estimation for the per-group effective channels $\boldsymbol{h}$ is given by
\begin{equation}
\boldsymbol{{\hat h}}
=\boldsymbol{\Theta}_{\rm s}^{-1}\boldsymbol{X}_{\rm t}^{-1}\boldsymbol{y}_{\rm t}=\boldsymbol{{h}}+\boldsymbol{{h}}_{\rm e},\label{Eq:ChanEsti}
\end{equation}
where $\boldsymbol{{h}}_{\rm e}\triangleq\boldsymbol{\Theta}_{\rm s}^{-1}\boldsymbol{X}_{\rm t}^{-1}\boldsymbol{z}_{\rm t}$  denotes the channel estimation error in $\boldsymbol{{\hat h}}$. As such, the MSE of the per-group effective channel estimation in each block is given by
\begin{align}\label{Eq:InterMSE}
{\rm MSE}(\boldsymbol{{\hat h}})&=\mathbb{E}\l[|| \boldsymbol{{h}}-\boldsymbol{{\hat h}}||
^2\r]=\mathbb{E}\l[||\boldsymbol{{h}}_{\rm e}||
^2\r]\nn\\
&=\mathbb{E}\l[\tr\l(\boldsymbol{\Theta}_{\rm s}^{-1}\boldsymbol{X}_{\rm t}^{-1}\boldsymbol{z}_{\rm t} \boldsymbol{z}_{\rm t}^{H}(\boldsymbol{X}_{\rm t}^{-1})^{H} (\boldsymbol{\Theta}^{-1}_{\rm s})^{H} \r)\r]\nn\\
&=\frac{\sigma^2}{P }\tr\l((\boldsymbol{\Theta}^{H}_{\rm s} \boldsymbol{\Theta}_{\rm s})^{-1}\r).
\end{align} 
\begin{algorithm}[t]
\caption{Proposed progressive channel estimation and passive beamforming designs.}
\label{Alg:Prog}
\begin{algorithmic}[1]
\STATE Initialize $i=1$.
\REPEAT
\STATE \textbf{Per-group effective  channel estimation}: Given $\boldsymbol{y}^{(i)}_{\rm t}\in\mathbb{C}^{M\times 1}$,   obtain the estimated per-group effective channels $\boldsymbol{{\hat h}}^{(i)}\in\mathbb{C}^{M\times 1}$ according to \eqref{Eq:IntraReciSig}, based on the designed basis training reflection matrix $\boldsymbol{\Theta}_{\rm s}$.
\STATE  \textbf{Intra-group channel estimation}: For each group $m$, first, collect the estimated per-group effective channels  over the first $i$ blocks, i.e., $\boldsymbol{{\hat \eta}}^{(i)}_m\in\mathbb{C}^{i\times 1}$.  Next,  resolve the  subgroup aggregated channels $\boldsymbol{{\hat g}}^{(i)}_m\in\mathbb{C}^{i\times 1}$  from $\boldsymbol{{\hat \eta}}^{(i)}_m$, based on the designed subgroup training reflection matrix $\boldsymbol{\Psi}^{(i)}_{{\rm{a}}}$.
\STATE  \textbf{Progressive passive beamforming}: Given the estimated subgroup aggregated channels of all groups, i.e., $\boldsymbol{{\hat g}}^{(i)}\in\mathbb{C}^{iM\times 1}$, optimize the passive beamforming vector $(\boldsymbol{\phi}^{(i)})^{H}\in\mathbb{F}^{iM \times 1}$ for rate maximization.
\STATE \textbf{Feedback and update}: The AP informs the IRS controller of $(\boldsymbol{\phi}^{(i)})^{H}$ for the phase-shift adjustment. Update $i=i+1$.
\UNTIL{$i >L$.
}
\end{algorithmic}
\end{algorithm}

Accounting for the IRS discrete phase shifts and the feasibility of the LS estimation, the optimization problem for minimizing the MSE of the per-group effective channel estimation can be formulated as follows. 
\begin{subequations}
\begin{align}
({\bf P1}):~~\min_{\boldsymbol{\Theta}_{\rm s}} ~~ &\frac{\sigma^2}{P }\tr\l((\boldsymbol{\Theta}^{H}_{\rm s} \boldsymbol{\Theta}_{\rm s})^{-1}\r) \nn \\  
~~~~\text{s.t.}~~~
& |[\Theta_{\rm s}]_{\tilde{m},m}|=1,~~~~ 1\le \tilde{m},m\le M,  \label{Eq:RefCons1}\\
& \angle[\Theta_{\rm s}]_{\tilde{m},m}\in\mathcal{F},~~~~ 1\le\tilde{m},m\le M, \label{Eq:RefCons2}\\
& \rank(\boldsymbol{\Theta}_{\rm s})=M, \label{Eq:RefCons3}
\end{align}
\end{subequations}
where  \eqref{Eq:RefCons1} and \eqref{Eq:RefCons2} respectively enforce the constraints of unit-modulus and discrete phase shift  on each entry of the basis training reflection matrix $\boldsymbol{\Theta}_{\rm s}$, and \eqref{Eq:RefCons3} guarantees the feasibility of the LS estimation.

\subsection{Proposed Basis Training Reflection Matrix Design}
First, it can be easily verified that problem (P1) is always feasible, since there exists a \emph{naive} basis training reflection matrix that satisfies all the constraints in \eqref{Eq:RefCons1}--\eqref{Eq:RefCons3},  regardless of the phase-shifter resolution and the pilot length. We denote it by $\boldsymbol{{\bar\Theta}}_{\rm s}$, whose entries are given by
\begin{align}\label{Eq:NaiveSch}
[\boldsymbol{{\bar\Theta}}_{\rm s}]_{{\tilde{m},m}}=\begin{cases}
-1,&~\tilde{m}=m,\\
1,&~\text{otherwise},\\
\end{cases}
\end{align}
where $\tilde{m}, m\in\cal{M}$.
However, despite its feasibility, the objective function of (P1) is non-convex due to the inverse operation as well as  the non-convex  constraints of full rank and unit-modulus. In addition, the phase shifts in the IRS basis training reflection matrix are restricted in a finite number of discrete values, rendering problem (P1) an NP-hard problem to solve. Numerically, the optimal solution to problem (P1) can be obtained by an exhaustive search over all possible basis training reflection matrices that satisfy the constraints in \eqref{Eq:RefCons1}--\eqref{Eq:RefCons3}, with the complexity order of  $\mathcal{O}(2^{bM^{2}})$, and then selecting the one with full rank and achieving the minimum MSE (MMSE). Note that the optimal solution may \emph{not be unique}. The computational complexity of the exhaustive search, however, may be practically prohibitive, since it increases exponentially with $M^2$ and $b$.

To address this issue, we first obtain the optimal solution to (P1) in some special cases of  $b$ and  $M$.  Then, for other cases in general, we propose a low-complexity algorithm to obtain a high-quality suboptimal  solution to problem (P1).
To this end, we first introduce two structured matrices: the DFT matrix and the Hadamard matrix.  Specifically, an $M\times M$ DFT matrix, denoted by $\bar{\boldsymbol{D}}_{M}$, is an orthogonal matrix whose entries are given by $[\bar{\boldsymbol{D}}_{M}]_{\tilde{m},m}=e^{-j\frac{2\pi (\tilde{m}-1) (m-1)}{M}}, 1\le {\tilde{m},m}\le M$.
On the other hand, a Hadamard matrix is also an orthogonal matrix while  its entries are either $+1$ or $-1$.  For example, a $4\times4$ Hadamard matrix, denoted by $\bar{\boldsymbol{H}}_{\rm 4}$, is given by
\vspace{-5pt}
\begin{equation}
\bar{\boldsymbol{H}}_4=\begin{bmatrix}
1  & 1 & 1& 1 \\
1  & -1 & 1& -1 \\
1  & 1 & -1& -1 \\
1 & -1 & -1& 1
\end{bmatrix}.\label{Eq:exaInv}
\end{equation}
Note that an $M \times M$ Hadamard matrix exists if and only if $M\in \mathcal{U}\triangleq\{u| u=2 ~\text{or}~u=4r, r\in\mathbb{Z}^{+}\}$.
The following proposition gives the optimal solution  to (P1) in two special cases.

\begin{proposition}\label{Lem:ContCE}\emph{For IRS with equally-spaced discrete phase shifts, the optimal solution to problem (P1) in the following two cases are given by:
\begin{itemize}
\item[1)] If $M\in \{2^{c}| c=1, 2, \cdots, b\}$, the DFT matrix $\bar{\boldsymbol{D}}_{M}$ is an optimal solution.
\item[2)] If $M\in \mathcal{U}$, the Hadamard matrix $\bar{\boldsymbol{H}}_{M}$ is an optimal solution.
\end{itemize}
}
\end{proposition}
\emph{Sketch of Proof:} First, it can be shown that if  there exists an orthogonal basis training reflection matrix, i.e., $\boldsymbol{\Theta}_{\rm s}^H \boldsymbol{\Theta}_{\rm s}=M\boldsymbol{I}$, satisfying all the constraints in \eqref{Eq:RefCons1}--\eqref{Eq:RefCons3}, then it is an optimal solution to problem (P1). Second, we can obtain the conditions under which  the orthogonal DFT and Hadamard matrices  satisfy the above constraints, thus completing the proof.
\hfill $\Box$

For other cases, in general, it is unknown whether there exists an orthogonal basis training reflection matrix satisfying all the constraints in \eqref{Eq:RefCons1}--\eqref{Eq:RefCons3}, which makes it hard to characterize the structure of the optimal solution to (P1). Thus we propose a \emph{novel} low-complexity  method, called \emph{DFT-Hadamard-based}  basis training reflection matrix design, to obtain a suboptimal  solution to problem (P1). Basically, our proposed design systematically constructs a \emph{near-orthogonal} basis training reflection matrix by performing the DFT-matrix quantization for $b\ge 2$, and the Hadamard-matrix truncation for $b=1$.
The rationalities and detailed construction are elaborated as follows.

\begin{itemize}
\item[1)] \textbf{DFT-based basis training reflection matrix for \bm{$b\ge 2$}}:  Our goal is to construct a quantized DFT matrix $\boldsymbol{D}_{M}$ for any $M$, such that it features \emph{near-orthogonality} in the sense that each entry has a value closest to that of the corresponding DFT matrix, but with the phase shift constrained in the feasible set $\mathcal{F}$.  Mathematically, we have  $\l[\boldsymbol{D}_{M}\r]_{\tilde{m},m}=e^{j{\check \theta}_{\tilde{m},m}}$, where ${\check \theta}_{\tilde{m},m}=\arg\min_{{\check \theta}_{\tilde{m},m}\in\mathcal{F}}\l| e^{j{\check \theta}_{\tilde{m},m}}- e^{-j\frac{2\pi (\tilde{m}-1) (m-1)}{M}}\r|.$
Such a quantized DFT matrix, however, can no longer preserve matrix invertibility for the IRS with any resolution of phase shifters. By extensive simulations, we observe that the quantized DFT matrix is always invertible for $b\ge2$ and achieves an MSE close to that of the continuous phase shifts when $b$ is sufficiently large. While, for the IRS with $1$-bit  phase shifters, i.e.,  $b=1$, the proposed quantized-DFT basis training reflection matrix, $\boldsymbol{D}_{M}$, is mostly \emph{noninvertible} for different $M$. For instance, we observe that  for $1\le M\le 20$,  $\boldsymbol{D}_{M}$ is invertible only when $M\in\{2, 4, 8, 16\}$, for which each of the quantized DFT matrices reduces to a Hadamard matrix with the same dimension. Thus, we resort to a Hadamard-based scheme as described below for designing the basis training reflection matrix when $b=1$.
\item[2)] \textbf{Hadamard-based basis training reflection matrix for \bm{$b=1$}}:  For the IRS with $1$-bit phase shifters,  by  leveraging the orthogonality of the Hadamard matrix, we propose to construct a \emph{truncated Hadamard} matrix for obtaining a \emph{near-orthogonal} basis training reflection matrix $\boldsymbol{H}_{M}$ as follows. First, find an $\ell\times \ell$ 
legitimate Hadamard matrix $\bar{\boldsymbol{H}}_{\ell}$ that has the smallest dimension $\ell$ while satisfying $\ell\ge M$. Then, truncate $\bar{\boldsymbol{H}}_\ell$ by preserving only the entries in the first $M$ rows and first $M$ columns. Mathematically, we have $[\boldsymbol{H}_{M}]_{\tilde{m},m}=[\bar{\boldsymbol{H}}_\ell]_{\tilde{m},m}, 1\le {\tilde{m},m}\le M.$
%
\end{itemize}
It is worth mentioning that  the optimal basis training reflection matrices in the special cases given in Proposition~\ref{Lem:ContCE} also comply with  the above proposed DFT-Hadamard-based design. Moreover, note that  the MSE of the proposed scheme is dependent on the designed  basis training reflection matrix  due to its \emph{non-orthogonality} in general, which is in sharp contrast to the case with continuous phase shifts for which the MMSE is shown to be a constant  given by $\sigma^2/P$ \cite{zheng2019intelligent}.

\section{Progressive Intra-Group Channel Estimation}\label{Sec:Intra-CE}
In this section, we detail our design for the intra-group channel estimation with the resolved per-group effective channels over the blocks. First, we  present how to design the subgroup partition over the blocks as well as the subgroup training reflection matrix for each block to resolve the subgroup aggregated channels for different groups. Next, we derive the MSE of the proposed intra-group channel estimation, by taking into account the channel estimation error due to both the per-group effective channel and intra-group channel estimations. 
\vspace{-8pt}
\subsection{Subgroup Partition and Training Reflection Matrix Design} \label{Sec:SubPart}

\begin{figure*}[t]
	\vspace{4pt}
	\begin{center}
		\includegraphics[height=4.5cm]{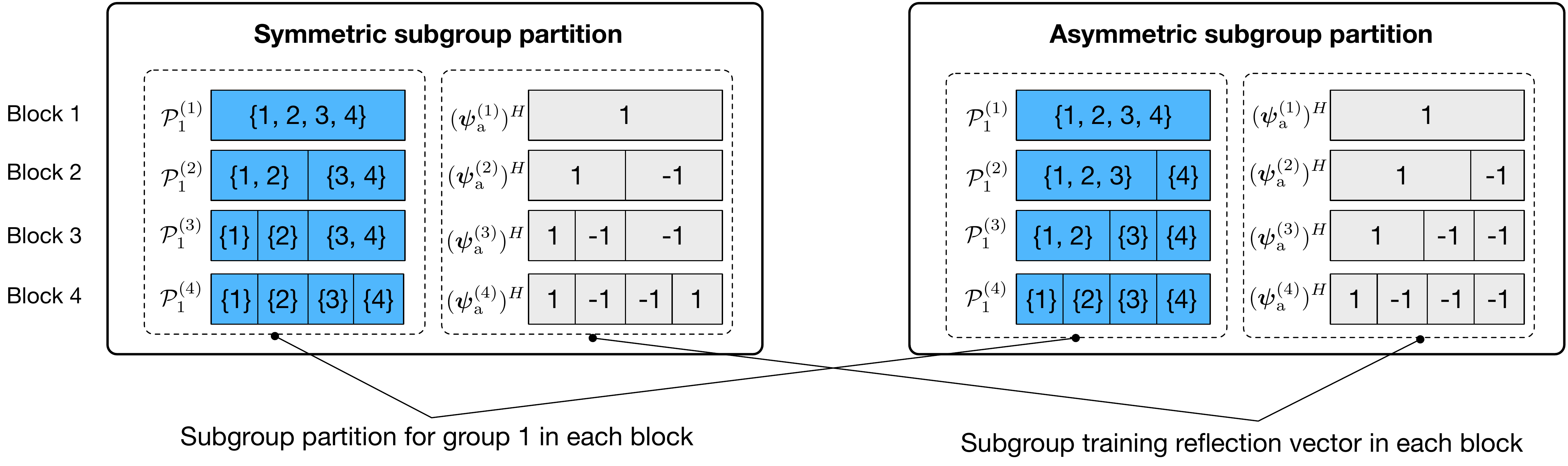}
		\caption{Proposed subgroup partition schemes and the corresponding subgroup training reflection matrix  design.}
		\label{FigIntraDiv}
	\end{center}
\end{figure*}

In each block $i$, as the intra-group channel estimation design applies to all groups, we consider a typical group $m$ for ease of elaboration. Let ${\cal P}^{(i)}_m=\{{\cal S}^{(i)}_{m,1},{\cal S}^{(i)}_{m,2},\cdots,{\cal S}^{(i)}_{m,i}\}$ denote the subgroup partition for group $m$ in block $i$, where ${\cal S}^{(i)}_{m,k}\ne \emptyset$ denotes the element-index set  for 
 subgroup $k$ with the indices arranged in an ascending order, and we have $\sum_{k=1}^{i}|{\cal S}^{(i)}_{m,k}|=L$ and $|{\cal P}^{(i)}_m|=i$, which implies that $i$ subgroup aggregated channels can be  resolved in block $i$.
 Recall that the subgroup training reflection vector in block $i$, $(\bm{\psi}_{\rm a}^{(i)})^{H}\in\mathbb{F}^{i\times 1}$,
is identical for all groups and all training symbols in each block, $g_{m,k}^{(i)}$  is the aggregated channel of the $k$-th subgroup with the elements indexed by ${\cal S}^{(i)}_{m,k}$, i.e., $g_{m,k}^{(i)}=\sum_{n\in {\cal S}^{(i)}_{m,k}} \underline{h}_n$, and $h^{(i)}_m$ is the effective channel of  group $m$ in block $i$, which is obtained by using the per-group effective channel estimation designed in Section \ref{Sec:Inter-CE}. 
Using the similar method as in Example~\ref{Exp:SubTrain}, by stacking $h^{(j)}_m$ with $j=1,2,\cdots,i$, we obtain
\vspace{-2pt}
\begin{align}\label{extraction}
\underbrace{\begin{bmatrix}
	h^{(1)}_m \\
	h^{(2)}_m \\
	\vdots \\
	h^{(i)}_m
	\end{bmatrix}}_{{\boldsymbol \eta}^{(i)}_m}
=
\underbrace{\begin{bmatrix}
	{\tilde {\boldsymbol \Psi}}_{\rm a}^{(i-1)} \\
	({\boldsymbol \psi}^{(i)}_{\rm{a}})^H
	\end{bmatrix}}_{{\boldsymbol \Psi}_{\rm a}^{(i)}}
~
\underbrace{\begin{bmatrix}
	g_{m,1}^{(i)} \\
	g_{m,2}^{(i)} \\
	\vdots \\
	g_{m,i}^{(i)}
	\end{bmatrix}}_{{\boldsymbol g}^{(i)}_m},
	\vspace{-8pt}
\end{align}
where 
${\boldsymbol \Psi}_{\rm a}^{(i)}\in\mathbb{F}^{i\times i}$ denotes the subgroup training reflection matrix in block $i$,
${\tilde {\boldsymbol \Psi}}_{\rm a}^{(i-1)}\in\mathbb{F}^{(i-1)\times i}$ denotes the \emph{extended} subgroup training reflection matrix 
for block $i-1$ with ${\tilde {\boldsymbol \Psi}}_{\rm a}^{(0)}\triangleq \emptyset$ (the detailed construction of ${\tilde {\boldsymbol \Psi}}_{\rm a}^{(i-1)}$ will be explained later),
${{\boldsymbol \eta}^{(i)}_m}$ denotes the stacked effective channels of group $m$ in the first $i$ blocks, and 
${{\boldsymbol g}^{(i)}_m}$ denotes the subgroup aggregated channels of group $m$ in block $i$. Without loss of generality, we assume that 
$i$ subgroup aggregated channels have been resolved in block $i, 1\le i\le L-1$, which requires  that $\rank({\boldsymbol \Psi}_{\rm a}^{(i)})=i$ as observed from \eqref{extraction}. Then, we  can focus on designing the intra-group channel estimation in block $i+1$ by properly designing the subgroup partition and training reflection matrix.

First, we determine the subgroup aggregated channels to be resolved in block $i+1$,  ${{\boldsymbol g}_m^{(i+1)}}$, by designing the subgroup partition in this block, given the subgroup partition  in block $i$, ${\cal S}^{(i)}_{m,k}$, and its corresponding subgroup aggregated channels ${\boldsymbol g}^{(i)}_m$. Specifically, to resolve $i+1$ subgroup aggregated channels, we select a typical \emph{parent} subgroup $k^*$ in block $i$ (which consists of more than one element) and partition it into two smaller \emph{children} subgroups in block $i+1$, which can be mathematically expressed as ${\cal S}^{(i)}_{m, k^*}={\cal S}^{(i+1)}_{m, k^*}\cup {\cal S}^{(i+1)}_{m, k^*+1}$. Other subgroups (except for subgroup $k^*$)  remain unchanged in block $i+1$, while the corresponding element-index sets are re-labeled as
\vspace{-5pt}
\begin{equation}
{\cal S}_{m,k}^{(i+1)}=\begin{cases}
{\cal S}_{m,k}^{(i)},& 1\le k \le k^*-1,\\
{\cal S}_{m,k-1}^{(i)},&k^*+2\le k \le i+1.\\
\end{cases}
\end{equation}
As such, the subgroup partition for block  $i+1$ is ${\cal P}^{(i+1)}_m=\{{\cal S}^{(i+1)}_{m,1},{\cal S}^{(i+1)}_{m,2},\cdots,{\cal S}^{(i+1)}_{m,i+1}\}$ and the corresponding subgroup aggregated channels  are given by ${{\boldsymbol g}_m^{(i+1)}}=[g_{m,1}^{(i+1)}, g_{m,2}^{(i+1)} ,\cdots g_{m,i+1}^{(i+1)}]^T\in\mathbb{C}^{(i+1)\times 1}$, where $g_{m,k}^{(i+1)}=\sum_{n\in {\cal S}^{(i+1)}_{m,k}} \underline{h}_n,$ $ 1\le k\le i+1$ satisfying $g_{m,k^*}^{(i+1)}+g_{m,k^*+1}^{(i+1)}=g_{m,k^*}^{(i)}$ due to the partition of the parent subgroup in block $i$.
In the following examples, we propose a systematic approach to construct two types of subgroup partitions over the blocks.

\begin{example}[Symmetric subgroup partition]\label{Exp:Sym}\emph{As illustrated in Fig.~\ref{FigIntraDiv}(left), in each block $i+1$, the symmetric subgroup partition scheme selects the subgroup with the largest size in block $i$ as the parent subgroup, i.e., $k^*=\arg \underset{k=1,\cdots,i}{\text{max}} \quad |{\cal S}^{(i)}_{m,k}|$ and partitions it into two (as close as possible) equal-size children subgroups, i.e.,  $$|{\cal S}^{(i+1)}_{m,k^*}|=\left\lceil |{\cal S}^{(i)}_{m,k^*}|/2\right\rceil,~~ |{\cal S}^{(i+1)}_{m,k^*+1}|=\left\lfloor |{\cal S}^{(i)}_{m,k^*}|/2\right\rfloor.$$}
\end{example}
\begin{example}[Asymmetric subgroup partition]\label{Exp:Asym}\emph{The asymmetric subgroup partition scheme, as illustrated in Fig.~\ref{FigIntraDiv}(right), partitions its largest subgroup with size $|{\cal S}^{(i)}_{m,k^*}|$ into two asymmetric-size  subgroups: one with $|{\cal S}^{(i+1)}_{m,k^*}|=|{\cal S}^{(i)}_{m,k^*}|-1$ elements and the other with $|{\cal S}^{(i+1)}_{m,k^*+1}|=1$  element.
}
\end{example}
Based on the subgroup aggregated channels in block $i+1$, ${{\boldsymbol g}_m^{(i+1)}}$, \eqref{extraction} can be updated as
\begin{align}\label{extraction2}
\underbrace{\begin{bmatrix}
	h^{(1)}_m \\
	h^{(2)}_m \\
	\vdots \\
	h^{(i+1)}_m
	\end{bmatrix}}_{{\boldsymbol \eta}^{(i+1)}_m}
=
\underbrace{\begin{bmatrix}
	{\tilde {\boldsymbol \Psi}_{\rm a}}^{(i)} \\
	({\boldsymbol \psi}_{\rm{a}}^{(i+1)})^H
	\end{bmatrix}}_{{\boldsymbol \Psi}_{\rm a}^{(i+1)}}
~
\underbrace{\begin{bmatrix}
	g_{m,1}^{(i+1)} \\
	g_{m,2}^{(i+1)} \\
	\vdots \\
	g_{m,i+1}^{(i+1)}
	\end{bmatrix}}_{{\boldsymbol g}^{(i+1)}_m},
\end{align}
where $({\boldsymbol \psi}^{(i+1)}_{\rm{a}})^H$ is the new subgroup training reflection vector designed for block $i+1$, and ${\tilde {\boldsymbol \Psi}}_{\rm a}^{(i)}\in\mathbb{F}^{i\times (i+1)}$ is the extended subgroup training reflection matrix  in block $i$, which is generated from
the subgroup training reflection matrix ${{\boldsymbol \Psi}}_{\rm a}^{(i)}$ by following a similar procedure as illustrated in  Example~\ref{Exp:SubTrain} that replicates the training reflection coefficients of the parent subgroup to those of  children (created) subgroups. Mathematically, ${\tilde {\boldsymbol \Psi}}_{\rm a}^{(i)}$ can be expressed as 
\begin{align}\label{mat_update1}
\!\!\!{\tilde {\boldsymbol \Psi}}_{\rm a}^{(i)}=\left[
\left[{{\boldsymbol \Psi}}_{\rm a}^{(i)}\right]_{:,1:k^*},~
\left[{{\boldsymbol \Psi}}_{\rm a}^{(i)}\right]_{:,k^*},~
\left[{{\boldsymbol \Psi}}_{\rm a}^{(i)}\right]_{:,k^*+2 : i+1}
\right].
\end{align}

Next, we  address how to design  the subgroup training reflection matrix for each block $i$, ${\boldsymbol \Psi}_{\rm a}^{(i)}, 1\le i\le I_0$. It can be observed from \eqref{extraction2} that in block $i+1$, to resolve ${{\boldsymbol g}^{(i+1)}_m}$ from ${{\boldsymbol \eta}^{(i+1)}_m}$, the subgroup training reflection matrix, ${\boldsymbol \Psi}_{\rm a}^{(i+1)}$, should be of full rank, which requires proper design of the training reflection vector in block $i+1$, $({\boldsymbol \psi}^{(i+1)}_{\rm{a}})^{H}$. Similar to the design for the basis training reflection matrix in Section~\ref{Sec:Inter-CE}, the optimal subgroup training reflection matrices for minimizing the intra-group channel estimation MSE  given any subgroup partition (as will be derived in the next subsection) need to be found via the exhaustive search. To reduce the complexity, we propose a simple yet efficient approach to systematically design a feasible subgroup training reflection matrix for each block, iteratively from block $i=1$ to $i=L$,  with only two-level phase shifts  (i.e., reflection coefficients are either $+1$ or $-1$). In particular, we can simply set $({\boldsymbol \psi}^{(1)}_{\rm{a}})^H=1$ and 
\begin{align}\label{mat_update2}
({\boldsymbol \psi}^{(i+1)}_{\rm{a}})^H&=\left[\left[{{\boldsymbol \Psi}}_{\rm a}^{(i)}\right]_{i,1:k^*},~
(\psi^{(i+1)}_{{\rm a},k^*+1})^{\dag},~
\left[{{\boldsymbol \Psi}}^{(i)}_{\rm a}\right]_{i,k^*+2:i+1} \right],\nn\\ &\qquad\qquad\qquad\qquad\qquad1\le i\le L-1,
\end{align}
where $(\psi^{(i+1)}_{{\rm a},k^*+1})^{\dag}=-[{{\boldsymbol \Psi}}^{(i)}_{\rm a}]_{i,k^*}$, to achieve the full rank of the subgroup training reflection matrices ${\boldsymbol \Psi}^{(i)}_{\rm a}$ for all blocks of $i$. Fig.~\ref{FigIntraDiv}  illustrates the designed subgroup training reflection vectors in different blocks for the two cases of symmetric and asymmetric subgroup partitions given in Example~\ref{Exp:Sym} and Example~\ref{Exp:Asym}. Taking the symmetric subgroup partition illustrated in Fig.~\ref{FigIntraDiv} as an example, the subgroup training reflection matrices for  blocks $1\le i\le 4$ are constructed as follows, all  of which can be observed to be of full-rank.
\begin{align}
{\boldsymbol \Psi}_{\rm a}^{(1)}&=1, {\boldsymbol \Psi}_{\rm a}^{(2)}=\begin{bmatrix}
   1  & 1 \\
   1  & -1 
  \end{bmatrix},
  {\boldsymbol \Psi}_{\rm a}^{(3)}=\begin{bmatrix}
   1  & 1 & 1 \\
   1  & 1 & -1 \\
   1  & -1 & -1 \\
  \end{bmatrix},\nn\\
   {\boldsymbol \Psi}_{\rm a}^{(4)}&=\begin{bmatrix}
   1  & 1 & 1& 1 \\
   1  & 1  & -1& -1 \\
   1  & -1 & -1& -1 \\
   1 & -1 & -1& 1
  \end{bmatrix}.
\end{align}

\vspace{-7pt}
\subsection{MSE of Intra-Group Channel Estimation}
Based on the subgroup training reflection matrices designed in the preceding subsection, in each block $i$, the LS estimation of the subgroup aggregated channels for each group $m$, $\boldsymbol{{\hat g}}^{(i)}_m\in\mathbb{C}^{i\times 1}$,  can be obtained as
\vspace{-5pt}
\begin{equation}
\boldsymbol{{\hat g}}^{(i)}_m= ({\boldsymbol \Psi}^{(i)}_{\rm a})^{-1}\boldsymbol{{ \hat \eta}}^{(i)}_m, ~~~\forall m\in\mathcal{M}.
\end{equation}
Recall that $\boldsymbol{{\hat g}}^{(i)}=[(\boldsymbol{{\hat g}}^{(i)}_1)^{H},  \cdots, (\boldsymbol{{\hat g}}^{(i)}_M)^{H}]^{H}\in\mathbb{C}^{iM\times1}$ denote the estimated subgroup aggregated channels of all groups in block $i$, whose size increases with $i$. In summary, $\boldsymbol{{\hat g}}^{(i)}$ is obtained by a succession of two operations, including the per-group effective channel and the intra-group channel estimations. Mathematically,  $\boldsymbol{{\hat g}}^{(i)}$ can be expressed as
\begin{align}\label{Eq:SubgroupChan}
\!\!\!\!\boldsymbol{{\hat g}}^{(i)}&
=\underbrace{\begin{bmatrix}
	({\boldsymbol \Psi}^{(i)}_{\rm a})^{-1} & &\\
	& \ddots& \\
	& & ({\boldsymbol \Psi}^{(i)}_{\rm a})^{-1}
	\end{bmatrix}}_{\boldsymbol{E}^{(i)}}\underbrace{\begin{bmatrix}
	\boldsymbol{{\hat \eta}}^{(i)}_1\\
	\vdots\\
	\boldsymbol{{\hat \eta}}^{(i)}_M
	\end{bmatrix}}_{\boldsymbol{{\hat \eta}}^{(i)}_{1:M}}\nn\\
	&=
\underbrace{\boldsymbol{E}^{(i)} \boldsymbol{\Pi}^{(i)}}_{\boldsymbol{F}^{(i)}}\underbrace{\begin{bmatrix}
	\boldsymbol{{\hat h}}^{(1)}\\
	\vdots\\
	\boldsymbol{{\hat h}}^{(i)}
	\end{bmatrix}}_{\boldsymbol{{\hat h}}^{(1:t)}}= \boldsymbol{F}^{(i)} \underbrace{\begin{bmatrix}
\boldsymbol{{h}}^{(1)}\\
\vdots\\
\boldsymbol{{h}}^{(i)}
\end{bmatrix}}_{\boldsymbol{{h}}^{(1:t)}}+\boldsymbol{F}^{(i)} \underbrace{\begin{bmatrix}
	\boldsymbol{{h}}^{(1)}_{\rm e}\\
	\vdots\\
	\boldsymbol{{h}}^{(i)}_{\rm e}
	\end{bmatrix}}_{\boldsymbol{h}_{\rm e}^{(1:i)}},
\end{align}
where $\boldsymbol{\Pi}^{(i)}$ is an $iM\times iM$  permutation matrix satisfying $\boldsymbol{{\hat \eta}}^{(i)}_{1:M}=\boldsymbol{\Pi}^{(i)}\boldsymbol{{\hat h}}^{(1:i)}$. 
{\color{black}Therefore, the MSE of the intra-group channel estimation in each block $i$ is derived as
\begin{align}
&{\rm MSE}(\boldsymbol{{\hat g}}^{(i)})=\mathbb{E}\l[|| \boldsymbol{{g}}^{(i)}-\boldsymbol{{\hat g}}^{(i)}||
^2\r]=\mathbb{E}\l[|| \boldsymbol{{g}}^{(i)}_{\rm e}||
^2\r]\nn\\
&=\mathbb{E}\l[||\boldsymbol{F}^{(i)}\boldsymbol{h}_{\rm e}^{(1:i)}||
^2\r]\nn\\
&  =\tr\l(\boldsymbol{F}^{(i)}\mathbb{E}\l[ \boldsymbol{h}_{\rm e}^{(1:i)} (\boldsymbol{h}_{\rm e}^{(1:i)})^{H}    \r](\boldsymbol{F}^{(i)})^{H}\r) 
\nn\\
&\overset{(a)}{=}\tr\l(\boldsymbol{F}^{(i)}\mathbb{E}\!\l[\diag\l\{\boldsymbol{h}^{(1)}_{\rm e} (\boldsymbol{h}^{(1)}_{\rm e})^H, \!\cdots,\! \boldsymbol{h}^{(i)}_{\rm e} (\boldsymbol{h}^{(i)}_{\rm e})^H\r\}\r]\!\!(\boldsymbol{F}^{(i)})^{H}\r)\nn\\
&\overset{(b)}{=}\tr\l(\boldsymbol{F}^{(i)} \mathbb{E}\l[\boldsymbol{I}_{i}\otimes (\boldsymbol{h}_{\rm e} \boldsymbol{h}_{\rm e}^H)\r](\boldsymbol{F}^{(i)})^{H}\r)\nn\\
&=\frac{\sigma^2}{P } \tr\l(\boldsymbol{E}^{(i)} \boldsymbol{\Pi}^{(i)} \l\{\boldsymbol{I}_{i}\otimes (\boldsymbol{\Theta}^{H}_{\rm s} \boldsymbol{\Theta}_{\rm s})^{-1}\r\}(\boldsymbol{\Pi}^{(i)})^{H}(\boldsymbol{E}^{(i)})^{H} \r) \nn\\
&\overset{(c)}{=}\frac{\sigma^2}{P } \tr\l(\boldsymbol{E}^{(i)} \l\{ (\boldsymbol{\Theta}^{H}_{\rm s} \boldsymbol{\Theta}_{\rm s})^{-1} \otimes  \boldsymbol{I}_{i}\r\}(\boldsymbol{E}^{(i)})^{H} \r)\nn\\
&\overset{(d)}{=}\frac{\sigma^2}{P }\sum_{m=1}^M \tr\l(({\boldsymbol \Psi}^{(i)}_{\rm a})^{-1}\zeta_m \boldsymbol{I}_{i}  \l(({\boldsymbol \Psi}^{(i)}_{\rm a})^{-1}\r)^H \r)\nn \\
&=\frac{\sigma^2}{P }\sum_{m=1}^M \zeta_m \tr\l(\l(({\boldsymbol \Psi}^{(i)}_{\rm a})^{H} {\boldsymbol \Psi}^{(i)}_{\rm a}\r)^{-1} \r)\nn\\
&=\frac{\sigma^2}{P } \tr\l((\boldsymbol{\Theta}^{H}_{\rm s} \boldsymbol{\Theta}_{\rm s})^{-1}\r) \tr\l(\l(({\boldsymbol \Psi}^{(i)}_{\rm a})^{H} {\boldsymbol \Psi}^{(i)}_{\rm a}\r)^{-1} \r),\label{Eq:MSEIntra}
\end{align} 
where $(a)$ holds since the per-group effective channel estimation errors over different blocks are independent; $(b)$ holds since the distribution of $\boldsymbol{h}^{(i)}_{\rm e}$ is the same for all blocks of $i$; $(c)$ is obtained from the property that for a permutation matrix $\boldsymbol{\Pi}^{(i)}$, we have  $$ \boldsymbol{\Pi}^{(i)} \l\{\boldsymbol{I}_{i}\otimes (\boldsymbol{\Theta}^{H}_{\rm s} \boldsymbol{\Theta}_{\rm s})^{-1}\r\}(\boldsymbol{\Pi}^{(i)})^{H}=(\boldsymbol{\Theta}^{H}_{\rm s} \boldsymbol{\Theta}_{\rm s})^{-1} \otimes  \boldsymbol{I}_{i};$$ $(d)$ holds since $\boldsymbol{E}^{(i)}$ is a block-diagonal matrix and $\zeta_m\triangleq[(\boldsymbol{\Theta}^{H}_{\rm s} \boldsymbol{\Theta}_{\rm s})^{-1}]_{m,m}$.   From \eqref{Eq:MSEIntra}, we can observe that the MSE of the intra-group channel estimation is determined by both the basis training reflection matrix $\boldsymbol{\Theta}_{\rm s}$ and the subgroup training reflection matrix ${\boldsymbol \Psi}^{(i)}_{\rm a}$ (except for $i=1$, where the derived MSE applies to the per-group effective channel estimation, which depends on the basis training reflection matrix only). Moreover, the MSE increases with $i$ due to the error accumulation and propagation arising from both the per-group effective channel and intra-group channel estimations.
}

\section{Progressive Passive Beamforming Optimization}\label{Sec:PBFOpt}
In this section, we optimize the progressive passive beamforming at the IRS in each block based on the estimated group/subgroup aggregated channels, for maximizing the achievable rate for data transmission by taking into account the channel estimation error. 
\vspace{-5pt}
\subsection{Problem Formulation}
Given the estimated group/subgroup aggregated channels $\boldsymbol{{\hat g}}^{(i)}\in\mathbb{C}^{iM\times 1}$, 
we define $\boldsymbol{{\hat G}}^{(i)}\triangleq\boldsymbol{{\hat g}}^{(i)}(\boldsymbol{{\hat g}}^{(i)})^{H}\in\mathbb{C}^{iM\times iM}$ and the channel estimation error covariance matrix as 
\begin{align}\label{Eq:ErrorCova}
\boldsymbol{{R}}^{(i)}&\triangleq\mathbb{E}[\boldsymbol{{g}}^{(i)}_{\rm e} (\boldsymbol{{ g}}^{(i)}_{\rm e})^{H}]\nn\\
&=
\frac{\sigma^2}{P } \underbrace{\tr\l((\boldsymbol{\Theta}^{H}_{\rm s} \boldsymbol{\Theta}_{\rm s})^{-1}\r) \tr\l(\l(({\boldsymbol \Psi}^{(i)}_{\rm a})^{H} {\boldsymbol \Psi}^{(i)}_{\rm a}\r)^{-1} \r)}_{\boldsymbol{{R}}^{(i)}_{\rm a}}.
\end{align}
Then the SINR in \eqref{Eq:SINR} can be rewritten as
\vspace{-5pt}
\begin{align}
\gamma(\boldsymbol{\phi}^{(i)})&=\frac{P (\boldsymbol{\phi}^{(i)})^{H}\boldsymbol{{\hat G}}^{(i)}\boldsymbol{\phi}^{(i)}}{\sigma^2\l((\boldsymbol{\phi}^{(i)})^{H}\boldsymbol{{R}}^{(i)}_{\rm a}\boldsymbol{\phi}^{(i)}+1\r)}, ~~~\forall i\in\mathcal{I}.\label{Eq:SINRNew}
\end{align} 
Note that $\boldsymbol{{R}}^{(i)}_{\rm a}$ defined in \eqref{Eq:ErrorCova} depends on both the basis training reflection matrix  $\boldsymbol{\Theta}_{\rm s}$ and the subgroup training reflection matrix ${\boldsymbol \Psi}^{(i)}_{\rm a}$. A closer observation reveals that the SINRs in different blocks have similar forms as shown in \eqref{Eq:SINRNew}. The main differences lie on the increasing size of the passive beamforming vector $(\boldsymbol{\phi}^{(i)})^{H}\in\mathbb{F}^{1 \times iM}$ with $i$ as well as the block-varying $\boldsymbol{{\hat G}}^{(i)}$ and $\boldsymbol{{R}}^{(i)}_{\rm a}$, which do not affect the optimization methods for designing the passive beamforming in different blocks. Thus we omit the block index (i.e., superscript $(i)$) in the sequel of this section  for notational brevity.  Accordingly, the optimization problem for maximizing the average achievable rate in \eqref{Eq:Rate} under the constraints of unit-modulus and discrete phase shifts is equivalent to that given below for the SINR maximization (by dropping the constant term $P /\sigma^2$).
\begin{subequations}
\begin{align}
({\bf P2}):~\max_{\boldsymbol{\phi}} ~~&\frac{\boldsymbol{\phi}^{H}\boldsymbol{{\hat G}}\boldsymbol{\phi}}{\boldsymbol{\phi}^{H}\boldsymbol{{R}}_{\rm a}\boldsymbol{\phi}+1}& \nn \\  
\text{s.t.}~~
& |\phi_{\ell}|=1, ~\quad \ell=1, 2, \cdots, iM,\label{Eq:P2Unit}\\
& \angle{\phi_{\ell}}\in\mathcal{F},  ~~\ell=1, 2, \cdots, iM.\label{Eq:P2Phase}
\end{align}
\end{subequations}

\vspace{-8pt}
\subsection{Proposed Algorithm for Problem (P2)}
{\color{black}Problem (P2) is a non-convex optimization problem due to the constrains of unit-modulus and discrete phase shifts. Since the discrete phase shifts are constrained in a finite set  $\mathcal{F}$,  the optimal solution can be obtained by  the exhaustive search, for which the complexity is of order $\mathcal{O}(2^{biM})$ since $\boldsymbol{\phi}^{H}\in \mathbb{F}^{1 \times iM}$, which increases exponentially with $biM$ as $i$ increases.  
To reduce the  complexity, we propose in this subsection an efficient \emph{successive refinement} algorithm to solve problem (P2) sub-optimally based on different initialization methods as follows.\footnote{\color{black} Note that the computational complexity and solution quality of the successive refinement algorithm are critically determined by the adopted initialization method.}}
\subsubsection{Initialization Methods}
For the progressive passive beamforming design, three initialization methods are applied  first for setting the initial passive beamforming vector in each block, followed by the proposed successive refinement algorithm for further refining the passive beamforming.\\
\textbf{\underline {SDR-based initialization}}:
The semidefinite relaxation (SDR)-based initialization optimizes the initial passive beamforming in each block $i$ by using SDR techniques.
Specifically,  we first relax the constraint of discrete phase shifts in \eqref{Eq:P2Phase} of problem (P2) and denote the resultant problem as problem (P3) given below. 
\vspace{-3pt}
\begin{subequations}
\begin{align}
({\bf P3}):~\max_{\boldsymbol{\phi}} ~~&\frac{\boldsymbol{\phi}^{H}\boldsymbol{{\hat G}}\boldsymbol{\phi}}{\boldsymbol{\phi}^{H}\boldsymbol{{R}}_{\rm a}\boldsymbol{\phi}+1}& \nn \\  
\text{s.t.}~~
& |\phi_{\ell}|=1, ~\quad \ell=1, 2, \cdots, iM.\label{Eq:P3Unit}
\end{align}
\end{subequations}
For this problem, 
we define $\boldsymbol{\Phi}\triangleq\boldsymbol{\phi}\boldsymbol{\phi}^{H}$, which satisfies $\boldsymbol{\Phi}\succeq\boldsymbol{0}$ and $\rank(\boldsymbol{\Phi})=1$. Then we have  $\boldsymbol{\phi}^{H}\boldsymbol{{\hat G}}\boldsymbol{\phi}=\tr(\boldsymbol{{\hat G}}\boldsymbol{\phi}\boldsymbol{\phi}^{H})=\tr(\boldsymbol{{\hat G}}\boldsymbol{\Phi})$ and $\boldsymbol{\phi}^{H}\boldsymbol{{R}}_{\rm a}\boldsymbol{\phi}=\tr(\boldsymbol{{R}}_{\rm a}\boldsymbol{\Phi})$.  By relaxing the non-convex rank-one constraint, problem (P3) is  transformed to
\vspace{-5pt}
\begin{subequations}
\begin{align}
({\bf P4}):~~\max_{\boldsymbol{\Phi}} ~~&\frac{\tr(\boldsymbol{{\hat G}}\boldsymbol{\Phi})}{\tr(\boldsymbol{{R}}_{\rm a}\boldsymbol{\Phi})+1}& \label{Eq:P4Obj} \\  
~~~~\text{s.t.}~~
& \boldsymbol{\Phi}\succeq\boldsymbol{0},\\
& [\boldsymbol{\Phi}]_{\ell,\ell}=1, ~~~\ell=1, 2, \cdots, iM.\label{Eq:P4Diag}
\end{align}
\end{subequations}
Problem (P4) is still non-convex since the objective function is non-convex over $\boldsymbol{\Phi}$. To address this issue, we apply the Charnes-Cooper transformation to reformulate problem (P4) \cite{charnes1962programming}. To be specific, we define
\vspace{-7pt}
\begin{align}
\boldsymbol{A}=\frac{\boldsymbol{\Phi}}{\tr(\boldsymbol{{R}}_{\rm a}\boldsymbol{\Phi})+1},~~\xi=\frac{1}{\tr(\boldsymbol{{R}}_{\rm a}\boldsymbol{\Phi})+1}.
\end{align}
As such, we have $\boldsymbol{\Phi}=\frac{\boldsymbol{A}}{\xi}$ and  $\tr(\boldsymbol{{R}}_{\rm a}\boldsymbol{A})+\xi=1$.
Consequently, problem (P4) is equivalent to the following problem.
\vspace{-7pt}
\begin{subequations}
\begin{align}
({\bf P5}):~~\max_{\boldsymbol{A}, \xi} ~~&\tr(\boldsymbol{{\hat G}}\boldsymbol{A})& \nn \\  
~~~~\text{s.t.}~~~
& \tr(\boldsymbol{{R}}_{\rm a}\boldsymbol{A})+\xi=1, ~~\boldsymbol{A}\succeq\boldsymbol{0},\nn\\
& [\boldsymbol{A}]_{\ell,\ell}=\xi, ~~~\ell=1, 2, \cdots, iM .\nn
\end{align}
\end{subequations}
Problem (P5) is a semidefinite programming (SDP) and hence its optimal solution, denoted by $\{\boldsymbol{A}^*,\xi^*\}$, can be obtained by using existing solvers such as CVX \cite{grant2008cvx}.  Then the optimal solution to problem (P4) is given by $\boldsymbol{\Phi^*}\!=\!\frac{\boldsymbol{A^*}}{\xi^*}$.  Since $\boldsymbol{\Phi^*}$, in general, may not be of rank-one, i.e., $\rank(\boldsymbol{\Phi}^*)\!\neq\! 1$, the optimal objective value of problem (P4) serves as an upper bound of problem (P3) only. In this case, the Gaussian randomization method can be used to obtain a feasible and high-quality suboptimal  solution to problem (P3) based on the higher-rank solution obtained by solving (P4) \cite{wu2019intelligent}, which is denoted by $\boldsymbol{{\tilde\phi}}$.

 \begin{figure*}[t!]
\centering
\subfigure[Normalized MSE of per-group effective channel \newline\indent \quad~ estimation vs. $M$.]{\label{FigMSE_simu3}
\includegraphics[height=4.7cm]{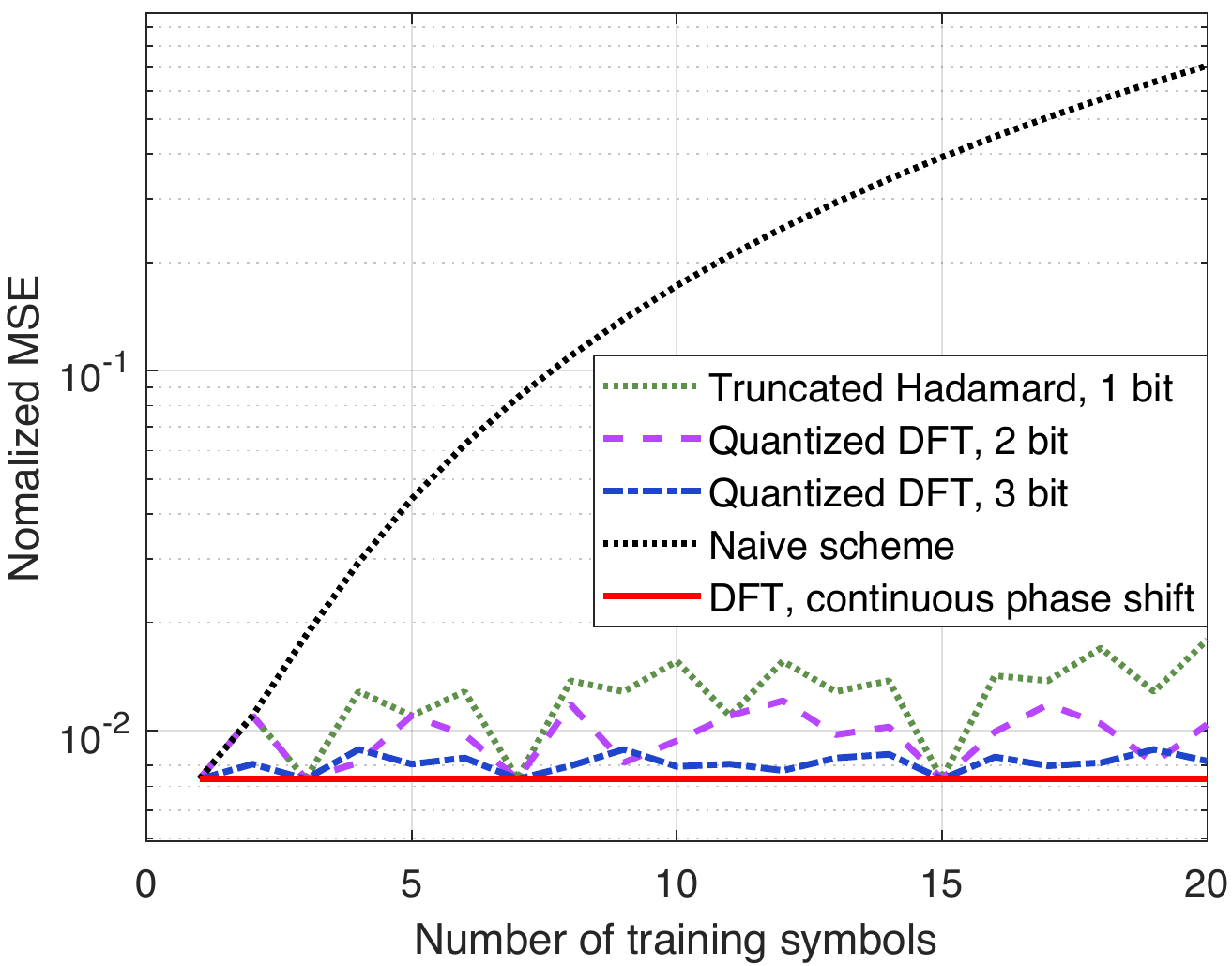}}
\hspace{25mm}
\subfigure[Achievable rate of passive beamforming  vs. $M$.]{\label{Fig1tierRate2_simu3}
\includegraphics[height=4.7cm]{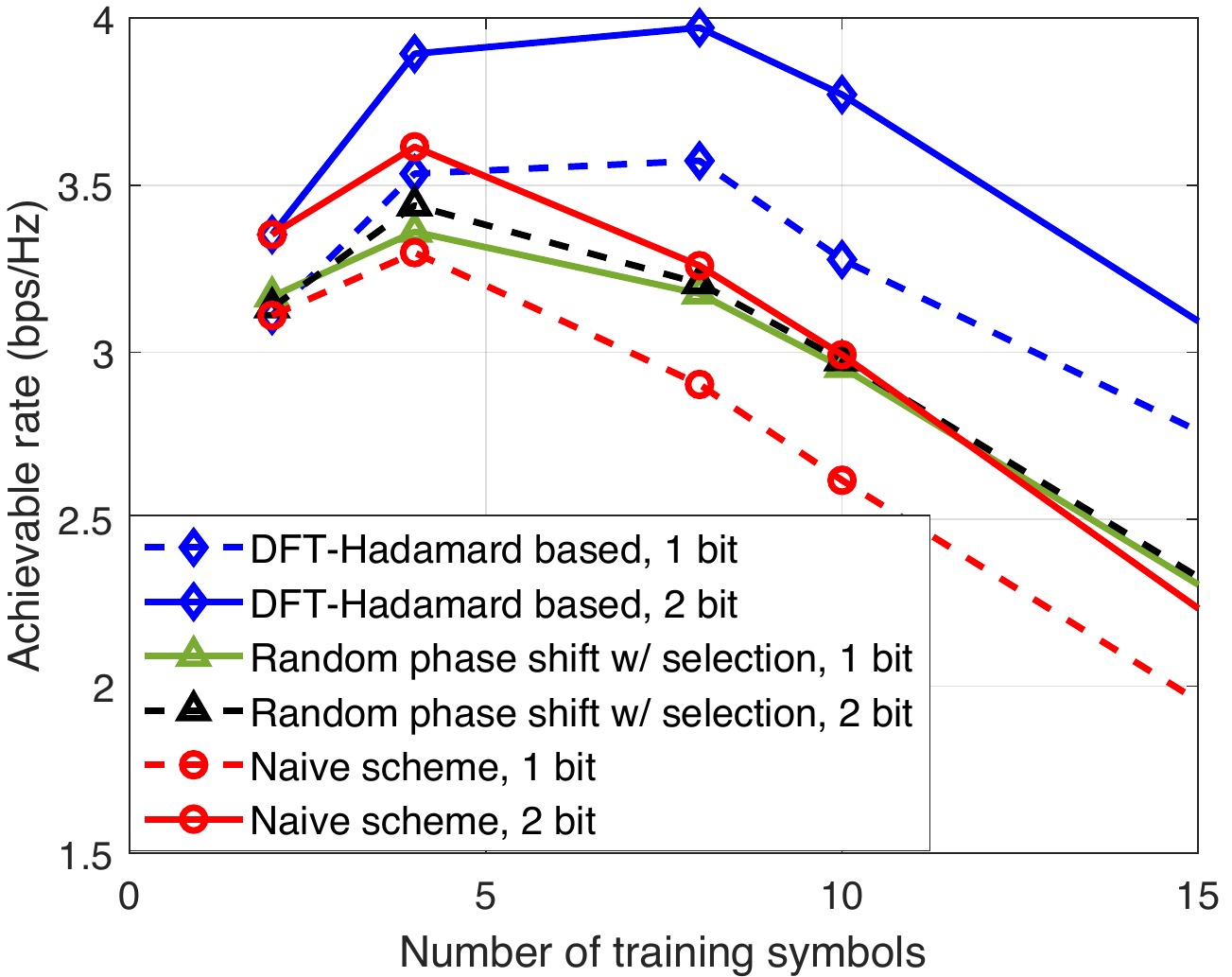}}
\caption{Performance  of the proposed per-group effective channel estimation and  corresponding passive beamforming design.}\label{Fig:TrajTime}
\end{figure*}

Next, based on the obtained near-optimal passive beamforming  $\boldsymbol{{\tilde\phi}}$ with continuous phase shifts, we  construct an initial IRS passive beamforming with discrete phase shifts by using \emph{phase quantization}. Specifically, 
for each of  ${\tilde\phi}_{\ell}, \ell\in\{1, \cdots, iM\}$, we directly quantize its phase shift to the nearest discrete value in $\mathcal{F}$, given by
\begin{equation}
\omega_{\ell}=\arg\min_{\omega_\ell\in\mathcal{F}}\l| e^{j\omega_\ell}- {\tilde\phi}_{\ell}\r|.
\end{equation}
{\color{black}Although the SDR-based initialization is expected to obtain a high-quality suboptimal solution to problem (P2), its complexity is in the order of $\mathcal{O}((iM)^{3.5})$ \cite{ben2001lectures}, which is practically affordable for moderate $i$ and/or $M$ but will be prohibitive when $iM$ becomes large.}\\
\textbf{\underline{Replication-based initialization}}: {\color{black}To reduce the complexity of the SDR-based initialization, we propose a replication-based initialization which makes use of the passive beamforming computed in the previous block for reducing the initialization complexity in the current block.} To this end, we set the initial passive beamforming in different blocks as follows. First, for block $i=1$, we compute the passive beamforming by the SDR-based initialization method followed by the successive refinement algorithm (to be described later). Then, in the subsequent blocks, we ``replicate" the passive beamforming in the previous block and augment it by adding the phase shifts of the children (created) subgroups, which are set as the same as those of their corresponding  parent subgroup (which is partitioned to create the new children sub-group, see Section~\ref{Sec:SubPart}). {\color{black}For block $i> 1$, the above  replication-based initialization has a negligible complexity in the order of $\mathcal{O}(1)$.}\\
\textbf{\underline{Channel-gain-maximization based initialization}}: {\color{black}Another approach to reduce the complexity for solving the SDP problem (P5) is to neglect the effects of the correlated channel estimation error (specified by $\boldsymbol{{R}}_{\rm a}$) and only maximize the channel power gain (i.e. $\boldsymbol{\phi}^{H}\boldsymbol{{\hat G}}\boldsymbol{\phi}$) instead of the SINR, referred to as the channel-gain-maximization based initialization.} In general, this problem is still NP-hard due to the constraint of discrete phase shifts (see \cite{wu2019beamforming}). To address this issue, we propose a simple yet efficient algorithm that first selects the strongest subgroup  that yields the largest channel power gain among the subgroup aggregated channels of all groups, and then tunes the discrete phase shifts of other subgroups to make their effective channels align with the strongest path in phase as close as possible. {\color{black}This initialization method has a low complexity order of  $\mathcal{O}(iM2^b)$ when $b$ is small.}
\subsubsection{Successive Refinement} Next, we  successively refine the passive beamforming based on the initialization for each block. Specifically, in each iteration, we find the optimal discrete phase shift for each subgroup to maximize the SINR in \eqref{Eq:SINRNew} via the one-dimensional search over $\mathcal{F}$, with those of the others being fixed, until the fractional decrease of $\gamma(\boldsymbol{\phi})$ in \eqref{Eq:SINRNew} is less than a sufficiently small threshold. The algorithm is guaranteed to converge since the objective value of (P2) is non-decreasing over the iterations and the optimal objective value of (P2) is upper-bounded by a finite value, i.e.,
\vspace{-5pt}
\begin{align}
\frac{\boldsymbol{\phi}^{H}\boldsymbol{{\hat G}}\boldsymbol{\phi}}{\boldsymbol{\phi}^{H}\boldsymbol{{R}}_{\rm a}\boldsymbol{\phi}+1}&=\frac{\boldsymbol{\phi}^{H}\boldsymbol{{\hat G}}\boldsymbol{\phi}}{\boldsymbol{\phi}^{H}(\boldsymbol{{R}}_{\rm a}+\frac{1}{M}\boldsymbol{I})\boldsymbol{\phi}}\nn\\
&=\boldsymbol{\phi}^{H}\boldsymbol{{U}}\boldsymbol{\phi}\le iM\lambda_{\max} (\boldsymbol{{X}}),
\end{align}
where $\boldsymbol{{U}}\triangleq(\boldsymbol{{R}}_{\rm a}+\frac{1}{M}\boldsymbol{I})^{-1}\boldsymbol{{\hat G}}$.
 {\color{black}Note that compared to the exhaustive search, our proposed successive refinement algorithm greatly reduces the complexity, which is in the order of $\mathcal{O}(\log(1/{\epsilon})iM2^b)$ for any feasible initialization, given the solution accuracy of $\epsilon>0$.}

\section{Numerical Results} \label{Sec:Num}
Numerical results are presented in this section to demonstrate the effectiveness of the proposed progressive channel estimation and passive beamforming designs. Unless specified otherwise, the system parameters are set as follows. Under the three-dimensional Cartesian coordinate system in meter (m), a single-antenna user located at $(20, 40, 0)$ transmits data to a single-antenna AP located at $(20, 0, 0)$. The IRS is equipped with a uniform rectangular array placed in the $y$--$z$ plane centered  at $(18, 30, 0)$, which consists of $N=80$ reflecting elements with half-wavelength spacing. 
For the large-scale fading, the distance-dependent path loss is modeled by $\beta(d)=\beta_0(d/d_0)^{-\alpha}$, where $d$ denotes the individual link distance, $\beta_0=-30$ dB denotes the reference channel gain at a distance of $d_0=1$ m, and ${\alpha}$ denotes the path loss exponent of the individual link and is set as $\alpha_{\rm UI} = 2.2$ and $\alpha_{
\rm IA} = 2.5$ for the user-IRS and IRS-AP links, respectively, by taking into account the longer distance of the latter than the former in practice. To account for the small-scale fading, we assume the Rician fading model for the associated channels with $K_{\rm UI}=3$ dB and $K_{\rm IA}=-20$ dB, respectively, which denote the Rician factors of the user-IRS and IRS-AP links. Moreover, each block consists of $30$ symbols.
Other parameters are set as $P=20$ dBm,  $\sigma^2=-89$ dBm,
 and $\Gamma=9$ dB.

\begin{figure*}[t]
\centering
\subfigure[Normalized MSE of channel estimation  vs.  \newline\indent \quad~ number of blocks.]{\label{FigMSE_20dB_group_1bit_simu3}
\includegraphics[height=4.6cm]{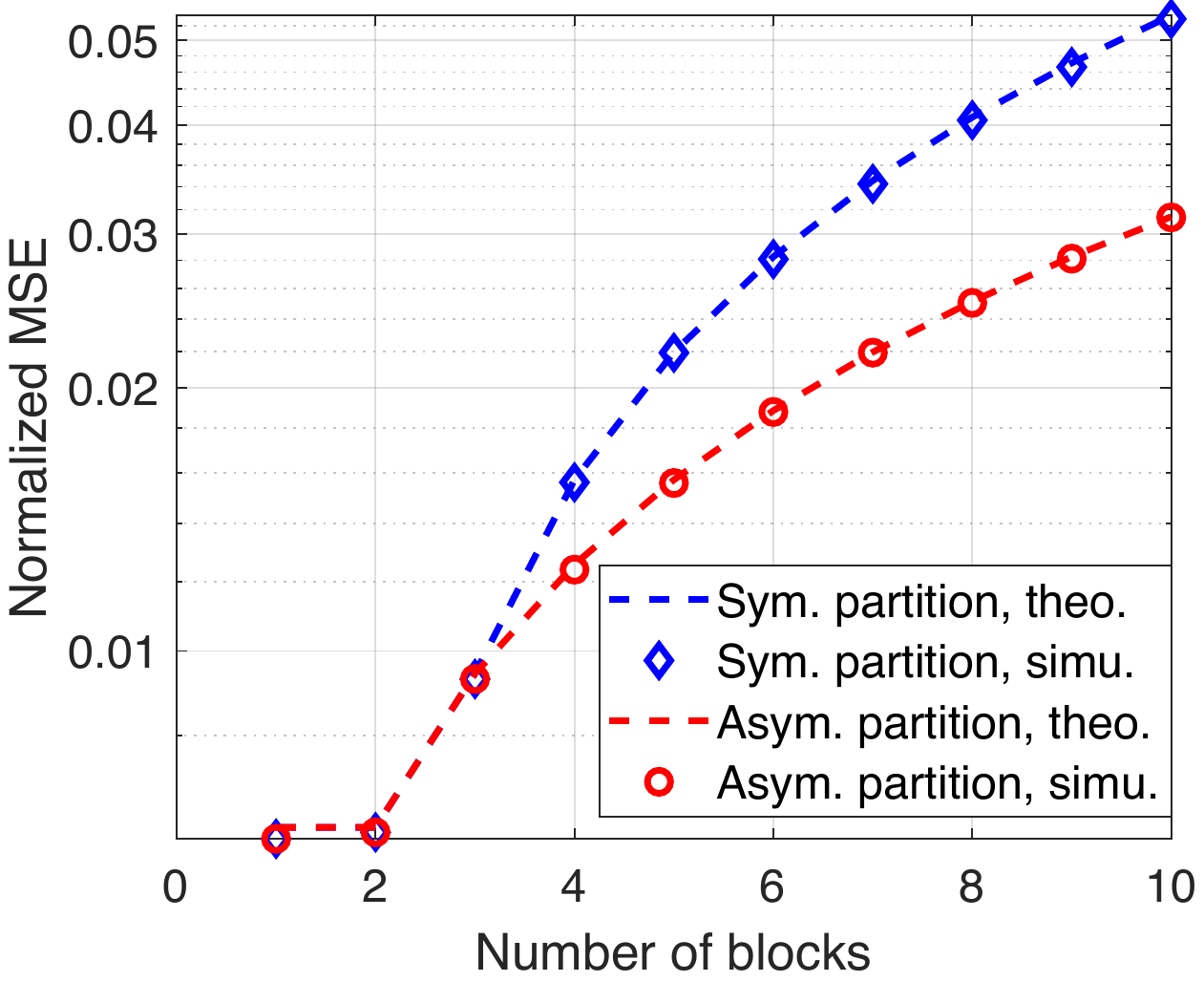}}
\subfigure[Achievable rate of passive  beamforming   vs. \newline\indent \quad~   number of blocks  with $M=4$.]{\label{FigMap80eleGroup4_group_1bit_simu3}
\includegraphics[height=4.6cm]{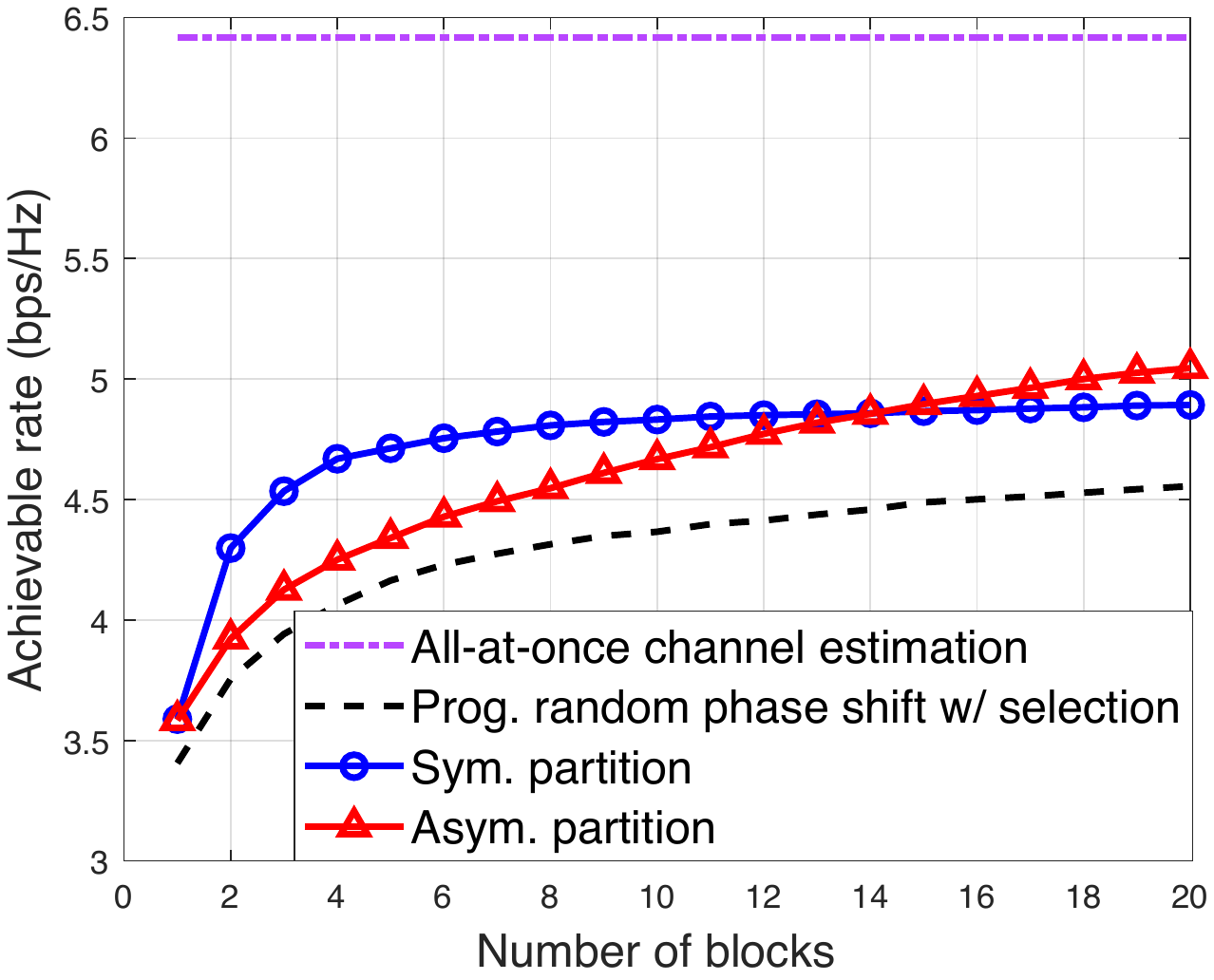}}
\subfigure[Achievable rate of passive beamforming vs.  \newline\indent \quad~ number of blocks  with $M=16$.]{\label{FigMap80eleGroup16_group_1bit_simu3}
\includegraphics[height=4.7cm]{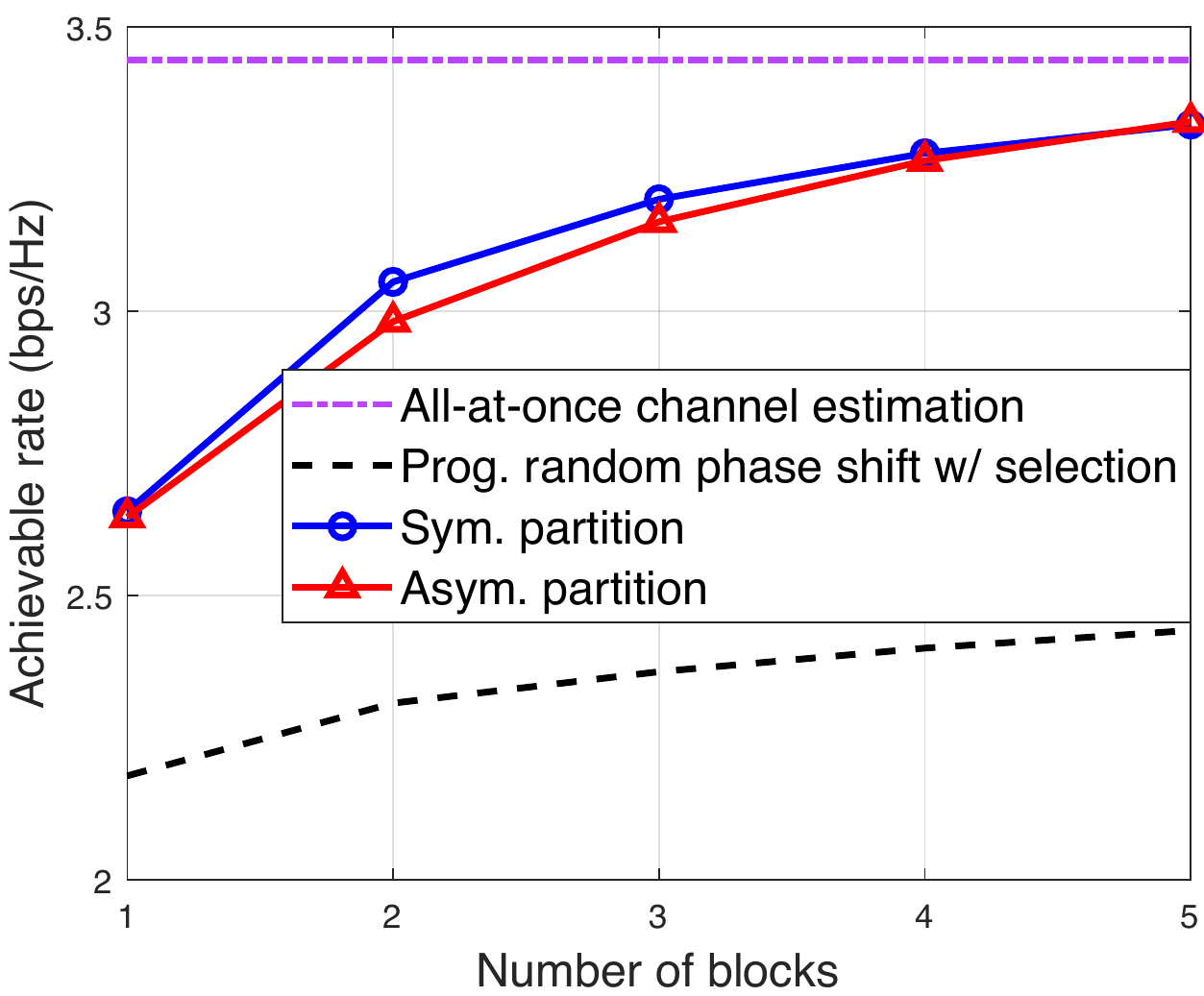}}
\label{Fig:TrajTime}
\caption{Performance  of the proposed intra-group channel estimation and corresponding passive beamforming design.}
\end{figure*}

\subsection{Per-Group Effective Channel Estimation}
First, we evaluate the performance  of the proposed DFT-Hadamard-based basis training reflection matrix design for estimating the per-group effective channel (thus, without loss of generality, we assume $I_0=1$ in this subsection) as well as the corresponding passive beamforming based on the estimated per-group effective channels.  We compare in Fig.~\ref{FigMSE_simu3} the normalized MSE of the per-group effective channel estimation 
 by the proposed basis training reflection matrix with that by  the naive scheme whose entries are given in \eqref{Eq:NaiveSch}. One can observe that the MSE of the naive scheme increases with the number of training symbols $M$  (or number of IRS groups),   since the corresponding basis training reflection matrix can be shown to be  more ill-conditioned. In contrast, our proposed design has much smaller MSE, especially for the IRS with high-resolution phase shifters (i.e., with larger value of $b$),   albeit that its MSE has small fluctuations  with $M$. Specifically, its MSE touches the lower bound with  continuous phase shifts when $M\in\{2, 4, 8, 16\}$, since the corresponding basis training reflection matrix reduces to a standard orthogonal Hadamard matrix.

In addition, we compare the rate performance of the proposed basis training reflection matrix design with two benchmark schemes: 1) naive scheme; 2) random phase shift with selection scheme: in each block $i$, the IRS generates $M$ sets of random phase shifts during the channel training and the AP selects the best set that achieves the largest passive beamforming gain among them.

To show the effect of $M$ on the achievable rate, 
 Fig.~\ref{Fig1tierRate2_simu3} plots the achievable rates of different schemes versus (vs.) $M$. First, it  is observed that there exists a tradeoff between the IRS channel estimation  and training overhead, since with too little training (i.e., smaller $M$)  the CSI is not accurate enough for achieving high  passive beamforming gain, while too much training results in less time for data transmission.
Second, our proposed design greatly outperforms the two benchmark schemes due to the properly designed  basis training reflection matrix.  In addition, one can observe significant rate improvement of the proposed design  by increasing the resolution of discrete phase shifters  from  $1$-bit  to $2$-bit, whereas the random phase shift with selection scheme shows marginal rate improvement only. Moreover, it is observed that without a properly designed basis training reflection matrix, the naive scheme for the IRS with $1$-bit phase shifters  performs  even worse than the random phase shift with selection  scheme.

\subsection{Intra-Group Channel Estimation}

Next, we evaluate the performance of the proposed subgroup training reflection matrices and the associated passive beamforming design.  In Fig.~\ref{FigMSE_20dB_group_1bit_simu3}, we  compare the normalized MSE of the proposed intra-group channel estimation with  the symmetric or asymmetric subgroup partition scheme vs. the number of blocks with $I_0=10$, assuming  the same DFT-Hadamard-based basis training reflection matrix design and the successive refinement algorithm with the replication-based initialization. First, it is observed that for both partition  schemes, the MSE results according to \eqref{Eq:MSEIntra}  match  well with the simulation results. Second,  the MSEs  of both partition schemes are non-decreasing with the increasing number of blocks due to the error accumulation and propagation. Besides, one interesting observation is that the asymmetric partition  yields smaller MSE than the symmetric counterpart in each block.

In addition,  we compare the achievable rate of the proposed passive beamforming design based on  the above  progressive channel estimation  with two benchmark schemes: 1) progressive random phase shift with selection scheme: which extends the previous random phase shift with selection scheme for single block to the case of  multiple blocks, i.e., for the second block, use $M$ more sets of random phase shifts, and select the best over them as well as those  in the first block, and so on; 2) all-at-once channel estimation:  the user  transmits  $M=N$ pilot symbols at one time for estimating all IRS elements' individual channels, which is equivalent to  the proposed per-group effective channel estimation with $M=N$ groups and  $L=N/M=1$ element per group; then, the passive beamforming is designed based on all channels estimated. It is worth noting  that the all-at-once channel  estimation design provides a performance upper bound on the achievable rate of the proposed progressive channel estimation design since it achieves the maximum  passive beamforming gain with fully resolved CSI and is free of the intra-group channel estimation error in the progressive channel estimation.

\begin{figure*}[t!]
\centering
\subfigure[Achievable rate vs. number of blocks.]{\label{FigBFGroup8Fram20Bit2_group_simu3}
\includegraphics[height=4.7cm]{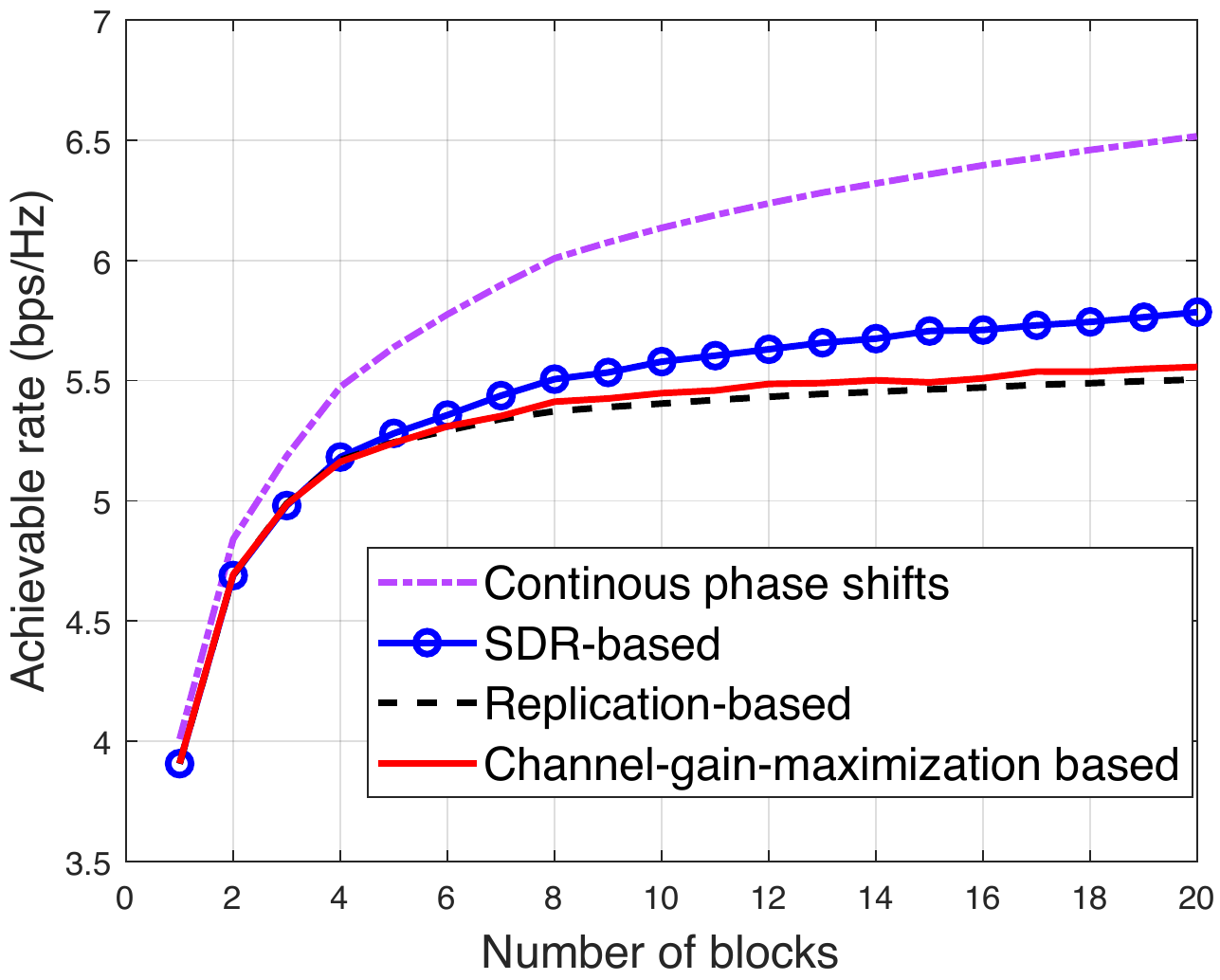}}
\hspace{25mm}
\subfigure[Effects of phase-shifter resolution.]{\label{FigBFGroup8Ele80_ele_dif_bit_simu3}
\includegraphics[height=4.7cm]{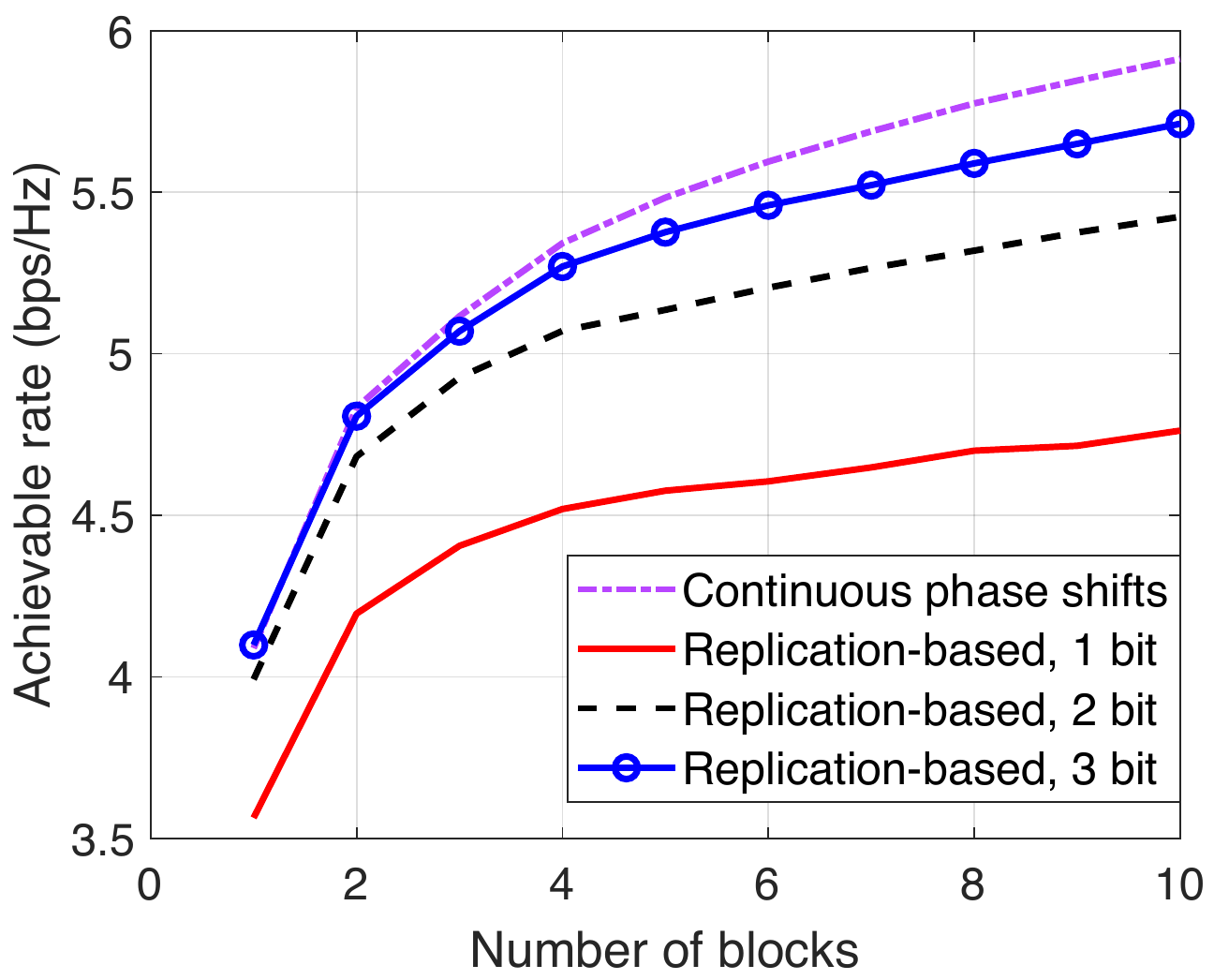}}
\caption{Performance  of the proposed passive beamforming design.}
\end{figure*}

   Figs.~\ref{FigMap80eleGroup4_group_1bit_simu3} and \ref{FigMap80eleGroup16_group_1bit_simu3} plot the achievable rate vs. the number of  blocks, with $b=1$ and $M=4$ vs. $M=16$. 
   Several important observations are made as follows. First, one can observe from Fig.~\ref{FigMap80eleGroup4_group_1bit_simu3} that although the proposed two subgroup partition schemes achieve almost the same rate in the first block, the rate of the symmetric partition grows  faster than the asymmetric counterpart in the subsequent blocks, but was finally overtaken by it after block $14$. This can be explained as follows. In the first few blocks, the symmetric partition tends to generate equal-size subgroups. This helps achieve more balanced estimation errors over the subgroups which in turn makes the passive beamforming more effective. 
    However, after sufficiently large number of  blocks, the subgroups generated by the asymmetric partition become similar to those by  the symmetric partition and its lower MSE (see Fig.~\ref{FigMSE_20dB_group_1bit_simu3}) helps achieve higher passive beamforming gain. Second, the fast rate convergence of the symmetric partition  indicates that, in practice, it is more suitable for the case with smaller $I_0$ as compared to the asymmetric partition. 
    Third, by comparing Figs.~\ref{FigMap80eleGroup4_group_1bit_simu3} and \ref{FigMap80eleGroup16_group_1bit_simu3}, we can observe that as $M$ increases, i.e., the number of training symbols per block increases or the number of IRS elements per group decreases, the achievable rate of the proposed progressive channel estimation and passive beamforming designs  approaches to that of the all-at-once channel estimation more closely as well as more quickly (in terms of the number of blocks). 
However, this may not be practically affordable as increasing $M$ leads to a larger number of pilot symbols per block as well as longer block length (assuming fixed training  overhead), which is not suitable for short-packet or delay-sensitive data transmissions.  Last, our proposed design based on the two partition schemes greatly outperform the progressive random phase shift  with selection scheme, which shows that progressive  CSI refinement is more effective than random reflection based selection for improving 
 the passive beamforming performance.

\subsection{Progressive Passive Beamforming}
Last, we show the rate  performance  of the proposed successive refinement algorithm with different initialization methods for the progressive passive beamforming. For comparison, we also consider the case of continuous phase shifts by solving a similar but modified problem of (P3) to provide a rate performance upper bound. 
The proposed DFT-Hadamard-based basis training reflection matrix is adopted for the per-group effective channel estimation, and the symmetric subgroup partition  is adopted for the intra-group channel estimation. In Fig.~\ref{FigBFGroup8Fram20Bit2_group_simu3}, we compare the achievable rates of different initialization methods  vs. the number of blocks with $M=4$ and $b=2$.
 It is observed that the low-complexity replication-based and channel-gain-maximization based initialization methods achieve close rates to the SDR-based initialization method that is of much higher complexity, but their rate performance loss in general increases with the number of blocks, i.e., when more CSI is resolved for the IRS. 
In addition, the impact of the phase-shifter resolution on the achievable rate  is evaluated in Fig.~\ref{FigBFGroup8Ele80_ele_dif_bit_simu3}, where the successive refinement algorithm is initialized by the replication-based method and $M=8$. We observe that the achievable rate of the proposed algorithm increases with higher-resolution phase-shifters, and achieves very close rate performance  to the case with continuous phase shifts when $b=3$.

\section{Conclusion}\label{Sec:Conc}
{\color{black}In this paper, for an IRS-aided single-user communication system with discrete phase shifts, we showed  that the corresponding IRS training reflection matrix design greatly differs from that with continuous phase shifts and the passive beamforming for data transmission should be optimized by taking into account the correlated channel estimation error due to discrete phase shifts. Moreover, for the practical scenario with insufficient number of training symbols in each block, 
 we  proposed a novel hierarchical training reflection design for progressively estimating the IRS  channels based on IRS-elements grouping and partition. 
 Given the resolved subgroup aggregated channels, we designed the progressive IRS passive beamforming to improve the achievable rate for data transmission over the blocks.
A low-complexity successive refinement algorithm with properly-designed initializations was proposed to obtain  high-quality suboptimal solutions. Last, numerical results demonstrated the effectiveness of our proposed channel estimation and passive beamforming designs with practical  discrete phase shifts.}

This work considered a basic and simplified setup to focus on investigating the proposed new design approach, while its results can be readily extended to more general cases such as multi-antenna AP/user, multiple users/IRSs, frequency-selective fading channels, imperfect IRS reflection model as well as correlated time/frequency channels with partial channel statistical knowledge, and so on.  In particular, for IRS-aided multiuser communications, the proposed progressive channel estimation and passive beamforming  designs can be jointly optimized  with multiuser  scheduling based on estimated channels,  which is an interesting as well as more challenging problem to solve in future work.

\vspace{-10pt}

\end{document}